%% file: martinez_arnaiz_fluxes_revised.tex
\documentclass[useAMS,usenatbib]{mn2e}
\DeclareMathAlphabet{\mathsc}{OT1}{cmr}{m}{sc}
\def\testbx{bx}%
\DeclareRobustCommand{\ion}[2]{%
\relax\ifmmode
\ifx\testbx\f@series
{\mathbf{#1\,\mathsc{#2}}}\else
{\mathrm{#1\,\mathsc{#2}}}\fi
\else\textup{#1\,{\mdseries\textsc{#2}}}%
\fi}
\usepackage{natbib}
\usepackage{graphicx}
\bibpunct{(}{)}{;}{a}{}{,}
\usepackage{graphicx}
\usepackage{lscape,longtable}
\usepackage{txfonts}
\usepackage{color}

   \title[]
{The effect of magnetic activity saturation in chromospheric flux--flux relationships}

   \author[R. Mart\'inez-Arn\'aiz et al.]{R. Mart\'inez-Arn\'aiz\thanks{email: rma@astrax.fis.ucm.es}, 
J. L\'opez-Santiago, I. Crespo-Chac\'on, and D. Montes
\thanks{Based on observations collected with the FEROS spectrograph at the  
2.2 m telescope at the European Organisation for Astronomical Research  
in the Southern Hemisphere, Chile (Programme: 074.D-0016(A))}\\
Departamento de Astrof\'{\i}sica, Facultad de Ciencias F\'{\i}sicas, Universidad Complutense de Madrid, E-28040 Madrid, Spain}
\begin{document}

\date{}

\pagerange{\pageref{firstpage}--\pageref{lastpage}} \pubyear{}

\maketitle

\label{firstpage}

\begin{abstract}
We present a homogeneous study of chromospheric and coronal flux--flux relationships
using a sample of 298 late-type dwarf active stars with spectral types F to M. The chromospheric 
lines were observed simultaneously in each star to avoid spread due to long term 
variability. Unlike other works, we subtract the basal chromospheric contribution in 
all the spectral lines studied. For the first time, we quantify the departure of dMe stars from 
the general relations. We show that dK and dKe stars also deviate from the general trend.
Studying the flux--colour diagrams we demonstrate that the stars deviating from the general 
relations are those with saturated X-ray emission and that those stars also present saturation
in the H$\alpha$ line. Using several age spectral indicators, we show that they are 
younger stars than those following the general relationships. 
The non-universality of flux--flux relationships found in this work should be taken into account when 
converting between fluxes in different chromospheric activity indicators. 
\end{abstract}


\section{Introduction}
\label{intro}

The study of stellar chromospheric and coronal emission is fundamental for the understanding 
of the outer layers of late-type stars. A number of problems can be addressed 
with activity based studies, such as the generation of stellar magnetic fields by dynamo 
processes or the nature of coronal heating and its relation with stellar activity.
Determining the magnetic activity level, per se, is also important in other 
research topics, such as accurately measuring radial velocity \citep[e.g.][]{1997ApJ...485..319S} 
or detecting transiting planets \citep[e.g.][]{1997ApJ...474..503H}. 
Another example that has interested the astronomic community for the past few years is 
the effect of magnetic activity on the formation and evolution of planets and
their atmospheres \citep[e.g.][]{2000ApJ...533L.151C} and, in a more general 
context, the evolution of overall stellar magnetic activity with age. 

When the (sometimes substantial) contribution from the 
acoustically driven basal atmosphere is subtracted from the observed emission, power law 
relationships are found between the excess flux in different activity indicators 
\citep[see][]{2000ssma.book.....S}. 
The first empirical flux--flux relationships compared fluxes in chromospheric and coronal lines 
with the aim of  studying the magnetic structure of stellar atmospheres 
\citep{1987A&A...172..111S,1991A&A...252..203R}. Subsequent studies analysed 
the relationship among different chromospheric activity indicators, such as 
\ion{Ca}{ii} H \& K lines and H$_{\alpha}$ \citep{1990ApJS...72..191S,
1990ApJS...74..891R,1995A&A...294..165M,1996ASPC..109..657M,1996A&A...312..221M,
2007A&A...469..309C} or the \ion{Ca}{ii} infrared triplet
\citep{1993MNRAS.262....1T,
2005ESASP.560..775L,2007A&A...466.1089B,2010A&A...520A..79M}. 
Basically, \citet{1995A&A...294..165M,1996ASPC..109..657M,1996A&A...312..221M} and 
\citet{1990ApJS...72..191S} included mostly binary (in many cases also evolved) stars. 
\citet{2005ESASP.560..775L,2007A&A...469..309C} 
and \citet{2010A&A...520A..79M} based their studies on single stars with spectral types F to K. 
It is important to mention that there is no standard technique to correct 
for the photospheric contribution among the mentioned studies. This makes the comparison 
between data acquired by different researches difficult. Besides, in many of the previous 
studies, the observations of the two activity indicators were not simultaneous. This fact introduces 
an important scatter in the flux--flux relationships due to the intrinsic variability of 
stellar magnetic activity.
To date, 
these power law relationships have been found to be independent of effective temperature or 
of stellar luminosity class for classes II--V (provided that stars beyond mid-M type are 
excluded). 

Some previous studies found that M-dwarfs presented a slight departure 
from the main flux--flux relationships in some spectral lines. In particular, this departure was 
found for emission-line M dwarfs when comparing chromospheric indicators with transition region ones 
\citep{1986A&A...154..185O,1989A&A...219..239R,2000ssma.book.....S} or with the coronal soft X-ray emission 
\citep{1987A&A...177..143S}. However, all the departures from the general flux--flux relationships 
found in these studies were i) restricted to 
dMe stars, ii) restricted to the comparison between the chromosphere and outer layers (transition region 
and corona). Besides, the fact that the coronal X-ray and transition region UV observations were not 
simultaneous to those of the chromosphere, introduced an important source of scatter due to the intrinsic 
variation of active stars \citep{1992A&A...258..432S}. Later, \citet{2005ESASP.560..775L} 
showed that some active late-K and M type stars deviated from classical H$\alpha$--\ion{Ca}{ii} 
flux--flux relationships. They tentatively identified them with flare stars.  
The sample of M-dwarfs in all these studies has been systematically biased towards emission-line stars. 
Therefore, these studies 
could not conclude whether this departure from the main flux--flux relationships was restricted to 
late-type stars with clear emission line characteristics or was common to a larger group of late 
K- and M-type stars.
The main objective of this work is to check which type of stars depart from the main flux--flux 
relationships when two chromospheric indicators are compared. 
In addition, we aim to properly quantify this departure using chromospheric fluxes that are appropriately 
corrected from the atmospheric basal contribution. By so doing we will provide unique information on the 
validity of the use of such relationships when trying to convert between fluxes in different lines.

For this research, we used high resolution optical echelle spectra. 
A noteworthy advantage of echelle spectra is that they cover a large
fraction of the optical spectrum simultaneously. Therefore, they allow a simultaneous
observation of all the activity indicator lines present in this spectral range,
avoiding the spread in the relations caused by temporal variability of activity levels.
This fact implies a significant improvement of the flux--flux
relationships with respect to those previously obtained, because most of the latter
were built by using activity diagnostics that were not measured simultaneously.
In addition, the chromospheric fluxes obtained from this method 
are corrected for the basal chromospheric emission.

Details on the technical information of the observations and
data reduction are given in Section 2. Section 3 describes the
analysis of the observations and the obtained excess emission 
equivalent widths, excess surface fluxes and X-ray luminosities 
and fluxes. Finally, Sections 4  and  5 are devoted, respectively, to the 
discussion of the results and conclusions of this work. 

\section{Data selection and reduction}
\label{section:data_selection}

The present study is based on high resolution echelle spectra. The 
total sample comprises 298 main-sequence, late-type  (spectral types F to M), single,  
active stars. We used data from \citet{2010A&A...514A..97L} (hereafter LS10) for 144 stars and 
data from \citet{2010A&A...520A..79M} (hereafter MA10) for 173\footnote{We note that we only used data for 
the stars classified as active.}. The former sample is mainly formed by main-sequence  stars 
but also includes some stars members of young associations and kinematic groups. The 16 binaries 
of that sample have not been included in our study. The MA10 sample is formed only by main-sequence  
single stars. Thus, all the active stars in this sample have been included in our study. We refer the reader to 
the mentioned works for a detailed explanation on the observing runs, telescopes 
and instruments they used as well as the characteristics of their 
spectra. We also note that there are 22 common stars between LS10 and MA10. 
After eliminating all binaries in LS10 and cross-correlating this sample with MA10, 
we obtained a total of 279 late-type  stars. We note that MA10 and LS10 samples 
are complementary in terms of emission levels: while the former includes mainly low activity stars, the 
latter is principally formed by young active stars. This fact is important bearing in mind that we aim to 
obtain precise flux--flux relationships and determine whether they hold for all spectral types and 
activity levels.

Given that 
previous studies showed a peculiar behaviour for some late-K and M stars \citep{2005ESASP.560..775L}, 
and that MA10 and LS10 samples only include a small number of these types of stars, 
 we considered it necessary to increase the number of 
these types of stars in the total sample. To complete the sample 
and increase the ratio of late-K and M stars, we obtained high resolution echelle 
spectra of 21 late-K and M stars, some of them well-known members of
young associations and moving groups. Bearing in mind that previous studies suggested that those stars 
deviating from the main flux--flux relationships were only those with emission features, the stars chosen to 
increase the sample are late-K and M type with and without such features. The chosen stars were also selected 
on the grounds of their known activity levels and youth. 
Only those stars with no signatures of accretion that could, eventually, affect 
chromospheric emission have been used (see Section \ref{sec_flux--flux}). 
The observations of this new sample of late-K and M stars were carried out at the
 European Southern Observatory, ESO (La Silla, Chile) in February 2005 with FEROS 
(Fiber-fed Extended Range Optical Spectrograph) linked to the Cassegrain focus of the
2.2 m telescope, with the CCD 2048 x 4096 (0.15 $\mu$m/pixel). This
configuration provides observations within the spectral range 3500--9200
\AA~with a resolution of 48000 (reciprocal dispersion ranging from 
0.03 to 0.09 \AA/pixel from the red to the blue region of the spectrum) 
in a total of 39 orders. This campaign will be referred as the FEROS05 
observing run hereafter. Preliminary results for some of the stars 
observed in this observing run are found in 
\citep{2007seadmg,2008cooldmg}. We note that there is one star in common 
with LS10 and one in common with MA10.

We used the reduction procedures in the IRAF\footnote{IRAF is 
distributed by the National Optical Astronomy Observatories, which are 
operated by the Association of Universities for Research in Astronomy, 
Inc., under cooperative agreement with the National Science Foundation.} 
packages and the standard method: bias and dark subtraction, 
flat-field division, cosmic rays correction, scattered light subtraction, 
and optimal extraction of the spectra. Th-Ar lamps were used to perform the 
 wavelength calibration. Finally, all the spectra were normalized 
by using a cubic spline polynomial fit to the observed continuum. 
We note that both the reduction process and data analysis 
 for the stars in FEROS05 observing run were 
completely analogous to those used by LS10 and 
MA10 in their respective works. This, together 
with the fact that the technique used to obtain chromospheric fluxes (see Section 
\ref{sub:spectralsubtraction}) was the 
same, ensures the compatibility of all the data used in this study.

\begin{figure}
\begin{center}
\includegraphics[bb=55 360 278 615,width=7cm, keepaspectratio]{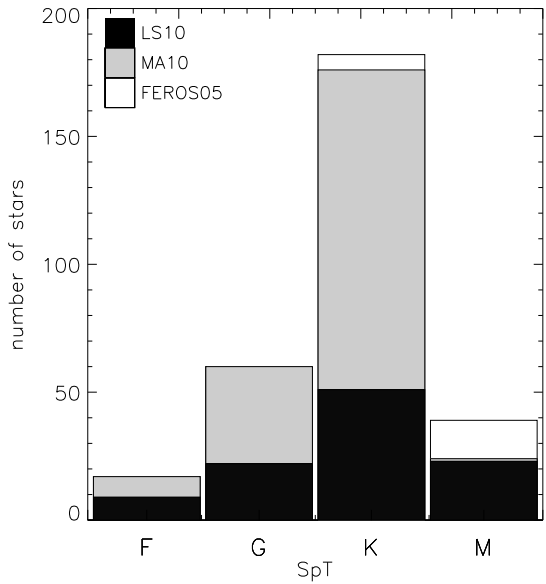}
\caption{Spectral type distribution of the complete stellar sample. Note that all the stars in the sample are 
active.}
\label{fig:histograma}
\end{center}
\end{figure}

The complete stellar sample contains 298 stars. 
The spectral type distribution 
 of the whole sample is: 17 F type stars, 60 G type stars, 182 K 
type stars and, 39 M type stars. Fig. 
\ref{fig:histograma} shows the spectral type distribution for LS10, MA10 and 
the FEROS05 observing run.
\section{Data analysis}

As we mentioned in Section~\ref{intro}, (power law) flux--flux relationships 
between chromospheric features are found only when basal chromospheric 
activity is subtracted.
This basal flux is common to active and non-active stars. Therefore, it is subtracted 
from the active star when using a non-active star as reference for the spectral 
subtraction instead of theoretical synthetic photospheric spectra (see MA10 and LS10 
for details). We used the same technique to reveal and measure equivalent widths of 
chromospheric lines in the FEROS05 stars. Then, we converted equivalent widths to
fluxes using empirical calibrations. In the following sections we give the details of the
analysis process.

\subsection{The spectral subtraction technique: Excess EW}
\label{sub:spectralsubtraction}
The spectral subtraction technique 
\citep[see $e.g.$][]{1995A&AS..114..287M,2000A&AS..146..103M}
is an unparalleled method to obtain the chromospheric 
contribution to the spectrum of a star. It permits the subtraction 
of the underlying photospheric contribution from the stellar spectrum. 
In this way, the spectral emission which originated at the chromosphere can be
studied in detail. When the spectrum of a non-active star with similar
spectral type, gravity and chemical composition is used, the basal chromospheric
flux is also subtracted (see MA10). Thus, any theoretical or semi-empirical 
calibration to subtract basal emission is avoided. As we mentioned above, 
this is important because power law relationships between 
the flux in different activity indicators only hold when the acoustically driven 
basal chromospheric emission is eliminated \citep{2000ssma.book.....S}.

For this work, we artificially constructed synthesized spectra using 
the program \textsc{jstarmod}\footnote{\textsc{jstarmod} is a modified
version of the Fortran code \textsc{starmod} developed at Penn State University 
\citep{1984BAAS...16..510H,1985ApJ...295..162B}. The modified code, implemented 
by J. L\'opez-Santiago, admits as input echelle spectra obtained with a CCD 
with more than 2048 pixels in the horizontal and/or vertical directions.}. 
The synthesized spectrum consists of the sum of 
rotationally broadened, \mbox{radial-velocity} shifted, and weighted spectra 
of non-active stars which are chosen to match the spectral types and luminosity 
classes of the components of the active star under consideration.
In this work, the non-active stars used as reference stars 
for the spectral subtraction were observed during the same 
observing run as their respective active stars. 
Once computed, the synthesized spectrum was subtracted from that of the target star 
to obtain a spectrum of the non-basal chromospheric contribution alone. Excess 
$EW$ of the activity indicators were measured in this subtracted spectrum. To estimate the errors 
in the measured $EW$, we followed \citet{2003A&A...411..489L} and considered
\textsc{jstarmod}'s typical internal precisions (0.5--2 km\,s$^{-1}$ in 
velocity shifts and $\pm$ 5 km\,s$^{-1}$ in projected rotational velocities, $v$~$\sin{i}$),
the $rms$ (root mean square) in regions outside the chromospheric features (typically 0.01--0.03), 
and the standard deviations. The estimated errors for relatively strong 
emitters are in the range 10\%--20\% but for low activity stars errors are larger. 
Taking into consideration that S/N is lower (higher $rms$) in the blue spectral region, 
errors in the chromospheric features at these wavelengths are larger. 
We refer the reader to LS10 and MA10 for a description of 
the stars used as references. In Table \ref{tab:ref_stars} we list the stars 
used as references for the FEROS05 stars.

\begin{table}
\caption{Stars used as reference to subtract the photospheric contribution to 
the spectrum of the FEROS05 stars.}
\label{tab:ref_stars}
\centering
\begin{tabular}{ l l l c c c  }
\hline
\noalign{\smallskip}
\multicolumn{1}{c}{Name} & \multicolumn{1}{c}{Other name} & \multicolumn{1}{c}{SpT} & 
\multicolumn{1}{c}{$B$-$V$} & v~$\sin{i}$\\ 
 &  &  & km s$^{\rm -1}$\\
\hline
\noalign{\smallskip}
HIP 83591 	&	HD 154363 & K5V	&	1.139	& 3.70$^{\rm 2}$ \\
HIP 62687 	&	HD 111631 & K7	&	1.348	& ... & \\
HIP 60661 	&	GJ 466        & M0V	&	1.153	& ... \\
HIP 25878 	&	HD 36395  & M1.5V	&	1.383	& 1.0$^{\rm 1}$ \\
HIP 51317 	&	GJ 393       & M2	&	1.448	& 1.1$^{\rm 1}$ \\
HIP 61706 	&     GJ 480       & M3	&	1.470	& 0.8$^{\rm 1}$ \\
HIP 36208 	&	GJ 273       & M3.5	&	1.438	& 0.0$^{\rm 1}$ \\
\noalign{\smallskip}
\hline
\multicolumn{2}{c}{$^{\rm 1}$ \citet{2009ApJ...704..975J}} &
\multicolumn{3}{c}{$^{\rm 2}$ \citet{2010A&A...514A..97L}}
\end{tabular}
\end{table}
\subsection{Excess surface fluxes}
\label{sec:fluxes}
Fluxes were derived for each of the activity indicators from its measured 
equivalent width ($EW_{\rm l}$) by using the continuum flux
\begin{equation}
F_{\rm s,l}=EW_{\rm l}\,F_{\rm s,l}^{\rm cont}\;\; \Longrightarrow\;\; 
\log F_{\rm s,l} = \log (EW_{\rm l})\,+\,\log(F_{\rm s,l}^{\rm cont}) ,
\end{equation}
where the continuum flux, $F_{\rm s,l}^{\rm cont}$, is dependent on the 
wavelength and must therefore be determined for the region where the activity indicator 
line is. We used the empirical relationships between $F_{\rm s,l}^{\rm cont}$ and
colour index, $B$--$V$, derived by \citet{1996PASP..108..313H} to compute 
$F_{\rm s,l}^{\rm cont}$ in the \ion{Ca}{ii} H \& K, H$\alpha$ and \ion{Ca}{ii} IRT regions. 
We note that the aforementioned relationships are linear for 
the spectral type range of our sample stars with the exception of the calibration for 
the H$\alpha$ region, for which \citet{1996PASP..108..313H} found small deviations from 
the linear trend for stars with $B$--$V$ $\ge$ 1.4. This implies 
that, in principle, the calibration for the H$\alpha$ region 
may not be valid for those stars cooler than M2/M3. 
We have explored the possibility of using the $\chi$-factor correction
\citep{2004PASP..116.1105W,2005PASP..117..706W,2008PASP..120.1161W}, which permits
the derivation of $L_{\rm H_{\alpha}}/L_{\rm bol}$ directly from the measured equivalent
widths of the emission lines and the colour index $B$--$V$ of the star by using
\begin{equation}
L_{\rm H_{\alpha}}/L_{\rm bol} = \chi \times EW_{\rm H\alpha}.
\end{equation}

\begin{figure}
\begin{center}
\includegraphics[width=8.5cm, keepaspectratio]{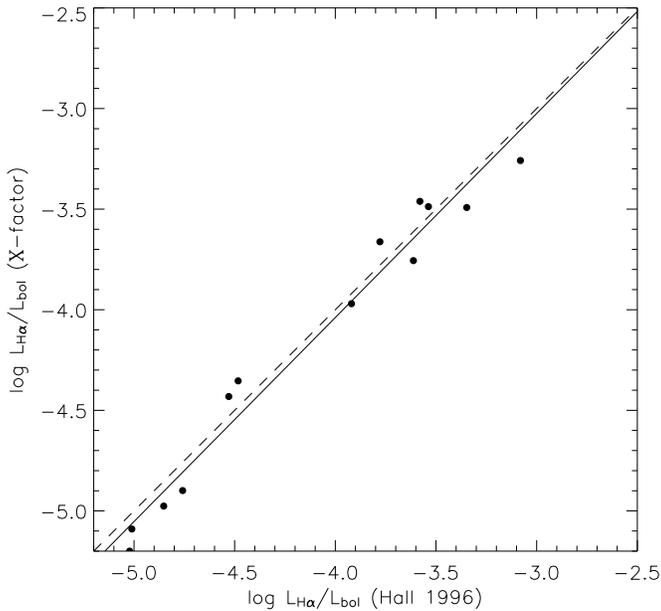}
\caption{Comparison between H$\alpha$ luminosities obtained using the $\chi$-factor 
correction and that in \citet{1996PASP..108..313H}. The dashed line is the 
one-to-one relation, while the continuous line is the linear fit to the data.}
\label{fig:chi_hall}
\end{center}
\end{figure}

To test whether the $\chi$-factor correction and that based on \citet{1996PASP..108..313H} 
calibration are consistent we have compared the values 
of $L_{\rm H_{\alpha}}/L_{\rm bol}$ obtained with both methods 
for M stars of the FEROS05 campaign (see Fig. \ref{fig:chi_hall}).  For these stars, 
bolometric luminosities ($L_{\rm bol}$) were determined using a ZAMS calibration.
The correlation between both results is clear: the linear regression 
equation is

\begin{equation}
(L_{\rm H_{\alpha}}/L_{\rm bol})_{\chi-factor} = 1.01\:(L_{\rm H_{\alpha}}/L_{\rm bol})_{Hall\; 1996} + 0.02\;(R~=~0.98).
\end{equation}

 Note that the maximum deviation in Fig. \ref{fig:chi_hall} is 0.15 dex, which is 
within the dispersion range in the relations found by \citet{1996PASP..108..313H} and those 
presented in \citet{2008PASP..120.1161W}. 
The fact that the obtained linear fit is compatible to the one-to-one 
implies that \citet{1996PASP..108..313H} calibrations for the continuum flux in the 
H$\alpha$ region can be used also for M0--M4 stars with the same level of confidence  as 
the $\chi$-factor correction. To ensure a homogeneous analysis of the data and obtain 
fluxes instead of  luminosities, we have used \citet{1996PASP..108..313H} calibration for 
all stars.

In Table \ref{tab:activity_flux}, we give the absolute flux at the stellar surface 
and its error for the late-K and M stars of the FEROS05 observing run. We refer the reader 
to LS10 and MA10 for a compilation of 
the fluxes of the rest of the stars.

\subsection{X-ray fluxes and luminosities}
\label{sec:xray}
X-ray flux is a direct measure of stellar activity because it is unlikely to include contributions 
from other sources such as the basal atmosphere \citep{1991A&A...252..203R,1992A&A...258..432S}. 
Therefore, in addition to the optical data, we searched for X-ray counterparts of the
stars in our sample in the ROSAT All Sky Survey (RASS) catalogue. For this purpose 
we followed \citet{2009A&A...499..129L}. We cross-correlated our total stellar sample with the 
ROSAT All-Sky Survey Bright Source Catalogue (RASS-BSC) and the Faint Source Catalogue 
(RASS-FSC) using a search radius of 30 arcsec to account for the ROSAT X-ray object 
coordinate determination accuracy. We found 243 counterparts for the sample of 298 stars.

To determine the X-ray fluxes, we used the count rate-to-energy flux
conversion factor ($CF$) relation found by \citet{1995ApJ...450..392S}

\begin{equation}
CF = (5.30 \cdot HR + 8.31) \cdot 10^{\rm -12}\; {\rm ergs}\,{\rm cm}^{\rm -2 }\,{\rm counts}^{\rm -1},
\end{equation}

where $HR$ is the hardness-ratio of the star in the ROSAT energy
band 0.1--2.4 KeV, defined as $HR = (H - S)/(H+S)$ with $H$ and $S$ being the
counts in the detector channels 11--49 and 52--201, respectively. 
X-ray fluxes were determined by multiplying the $CF$ value by the count-rate
\footnote{count-rate in the ROSAT energy band (0.1--2.4 KeV).}
of the sources in the same band. Note that the fluxes determined in this way
are observed fluxes, but not surface fluxes. Fluxes were then transformed 
into luminosities using the distances from the star to the Earth. 
Since the $CF$ and the count rate ($CR$) are defined for the ROSAT energy band
0.1--2.4 KeV,  the X-ray luminosity $L_{\rm X}$ is also integrated in this band.

To obtain surface X-ray fluxes we used the computed luminosities and the stellar radius. 
Since the stars in the sample are dwarfs (see LS10, MA10 and Table 
\ref{tab:parameters}), the radius were estimated by using  a ZAMS calibration.

The number of matches between our sample and the 
RASS decreases with increasing distance. 
For instance, \citet{2009A&A...499..129L} observed that, for their sample, at 
$d \le 40$\,pc, approximately 70\% of the stars were cross-identified, while 
at $d \le 50$\,pc, only half of them had an RASS counterpart. This suggests 
that some X-ray emitters at large distances are lost as a consequence of the flux 
limit of the RASS, producing a bias in our X-ray sample. Note
that the MA10 sample is limited to 25 pc and this effect should be negligible.

\begin{figure*}
\centering
\includegraphics[width=8cm, keepaspectratio]{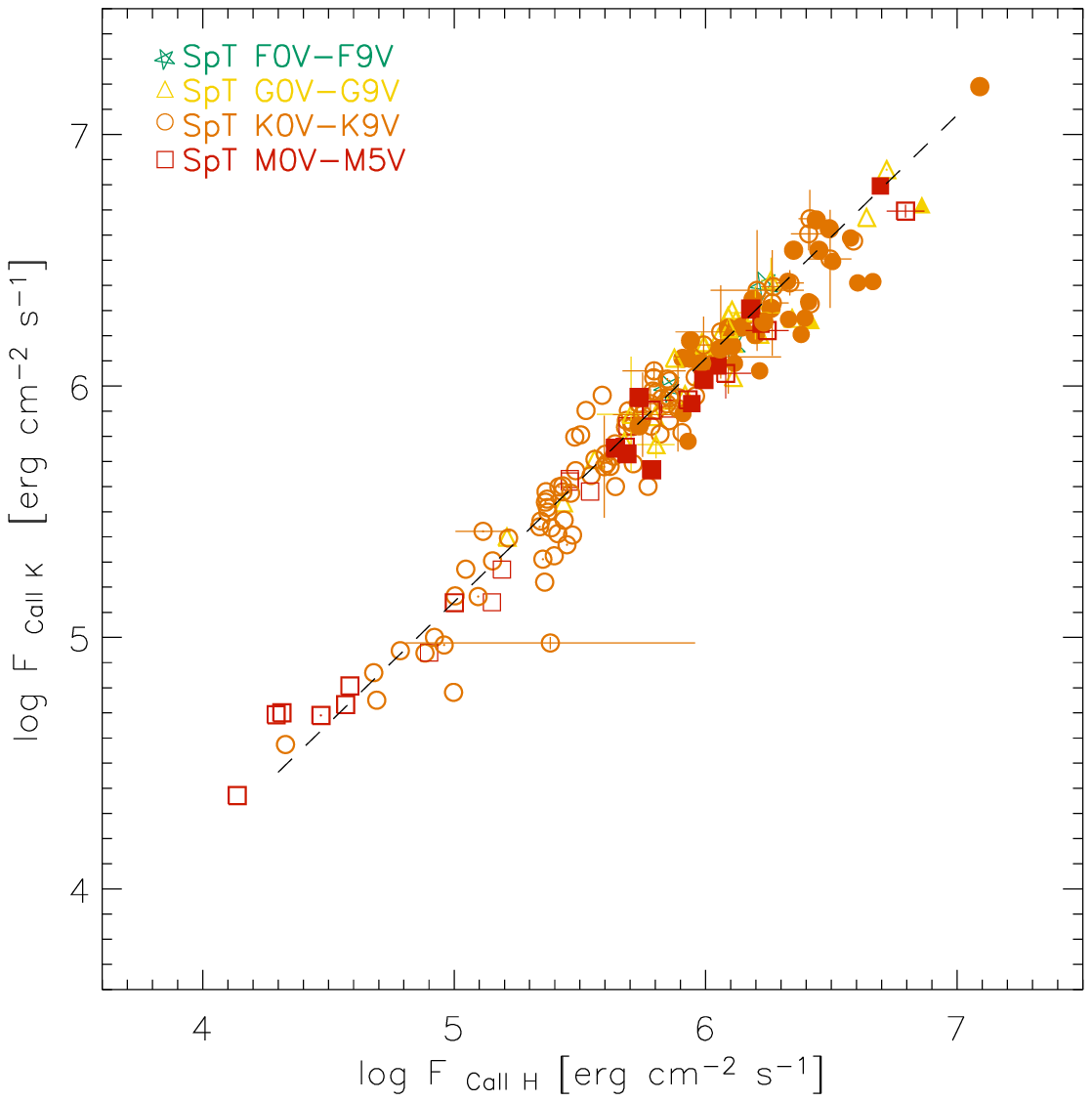}
\includegraphics[width=8cm, keepaspectratio]{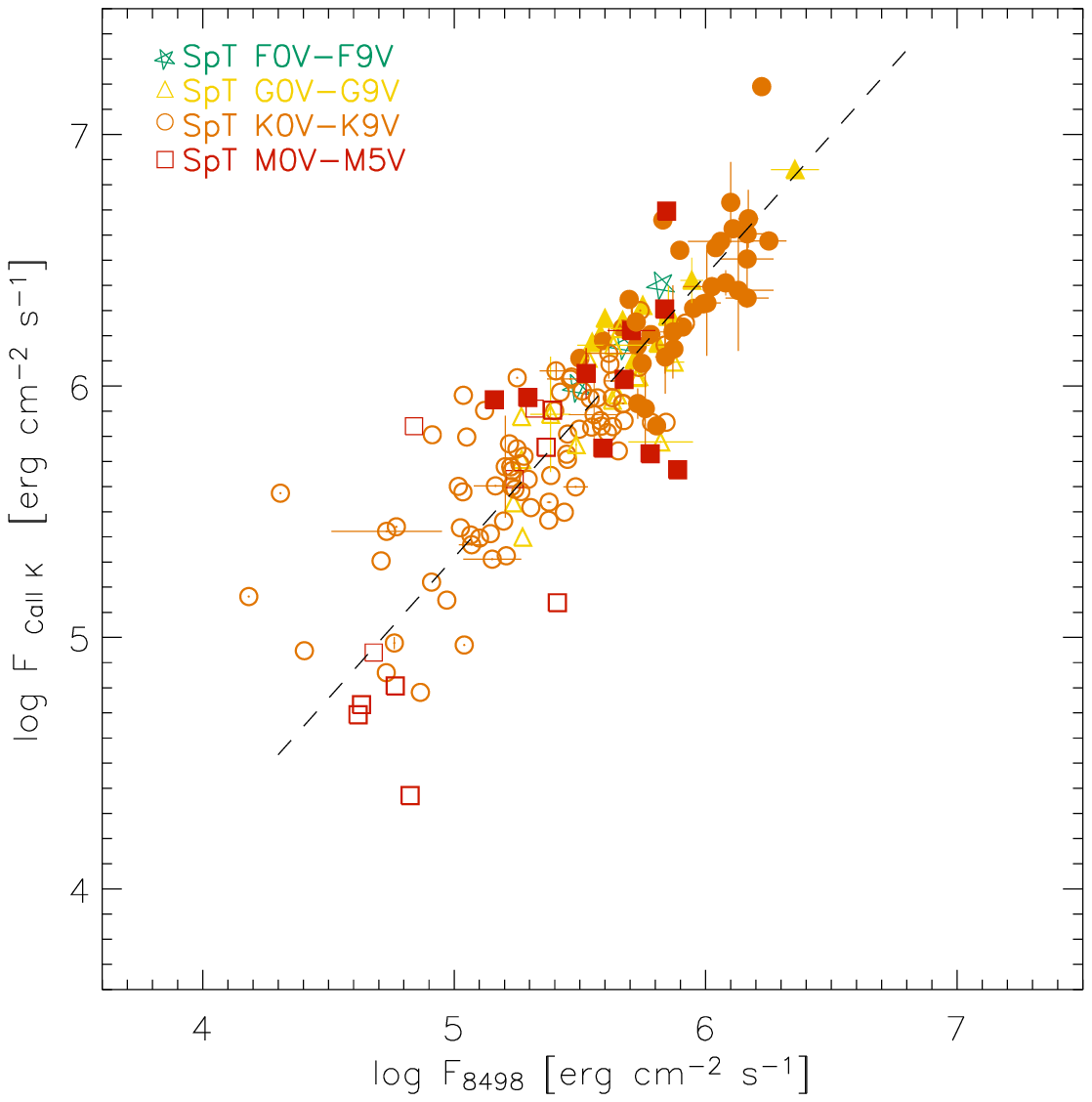}
\includegraphics[width=8cm, keepaspectratio]{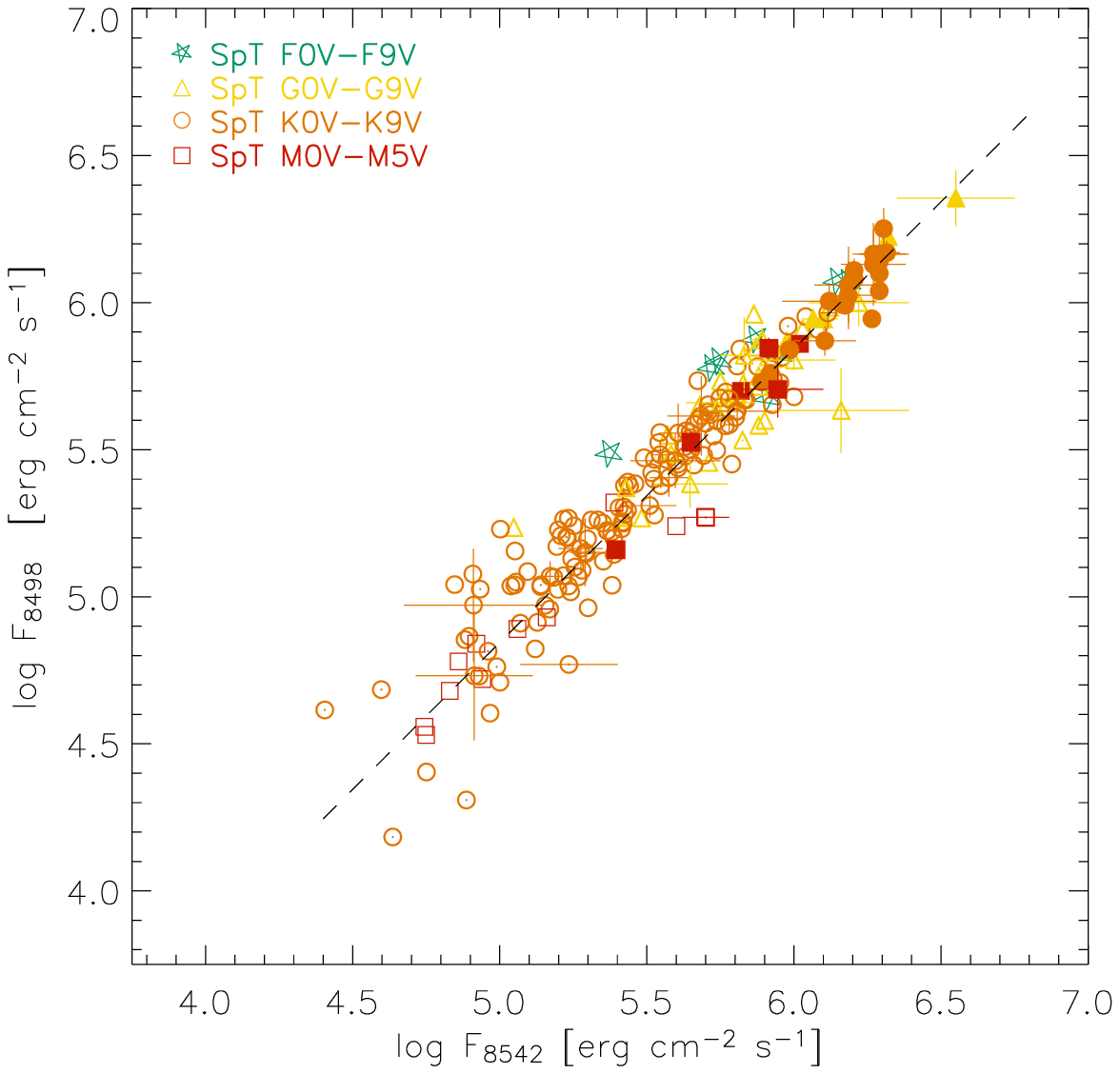}
\includegraphics[width=8cm, keepaspectratio]{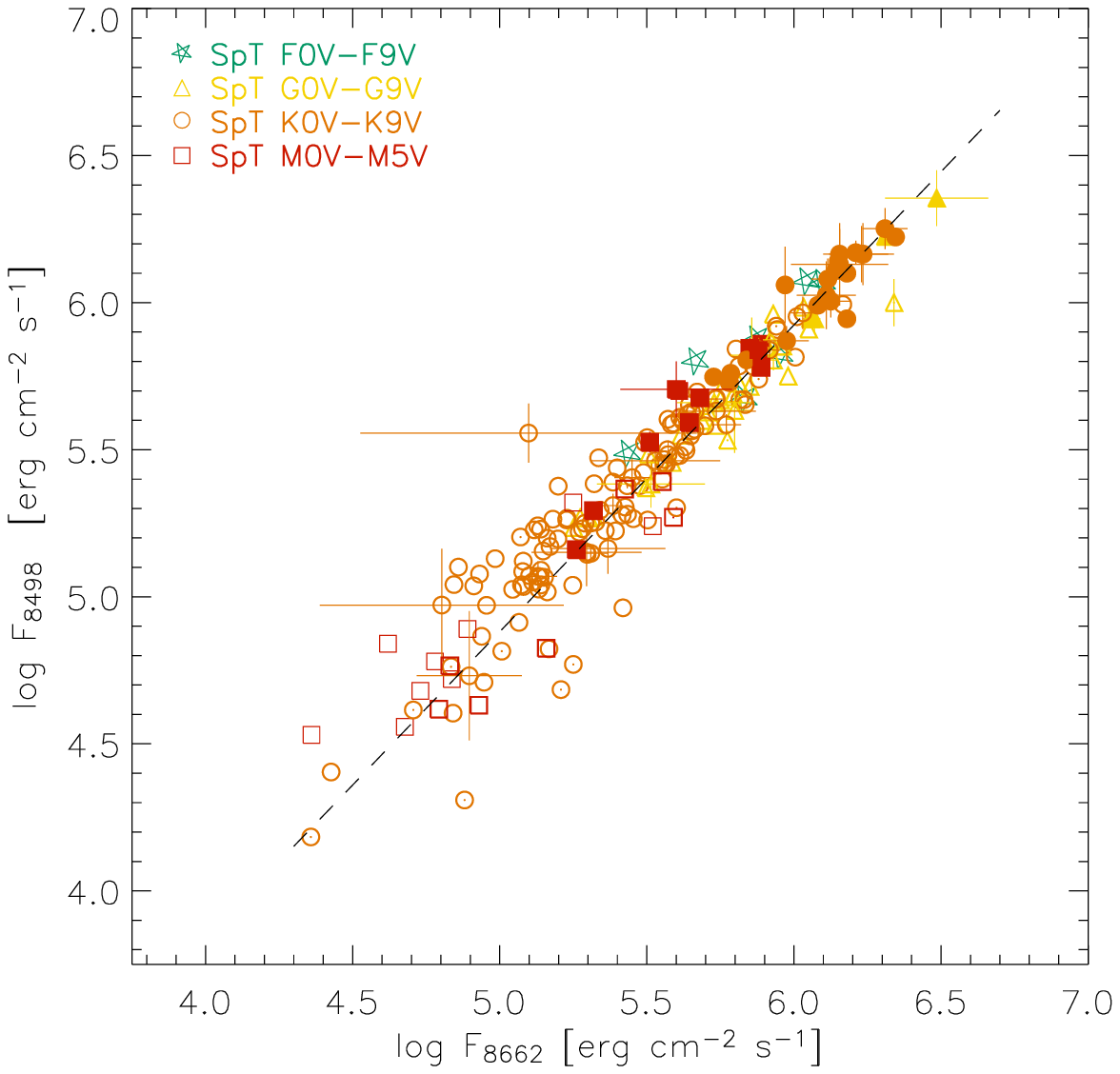}
\caption[]{Flux--flux relationships between calcium lines (\ion{Ca}{ii} H \& K and 
\ion{Ca}{ii} IRT). Green stars represent F type stars, yellow triangles G type stars, 
orange circles K type stars and red squares M type stars.}
\label{fig:flux_all}
\end{figure*}
\begin{figure*}
\centering
\includegraphics[width=8cm, keepaspectratio]{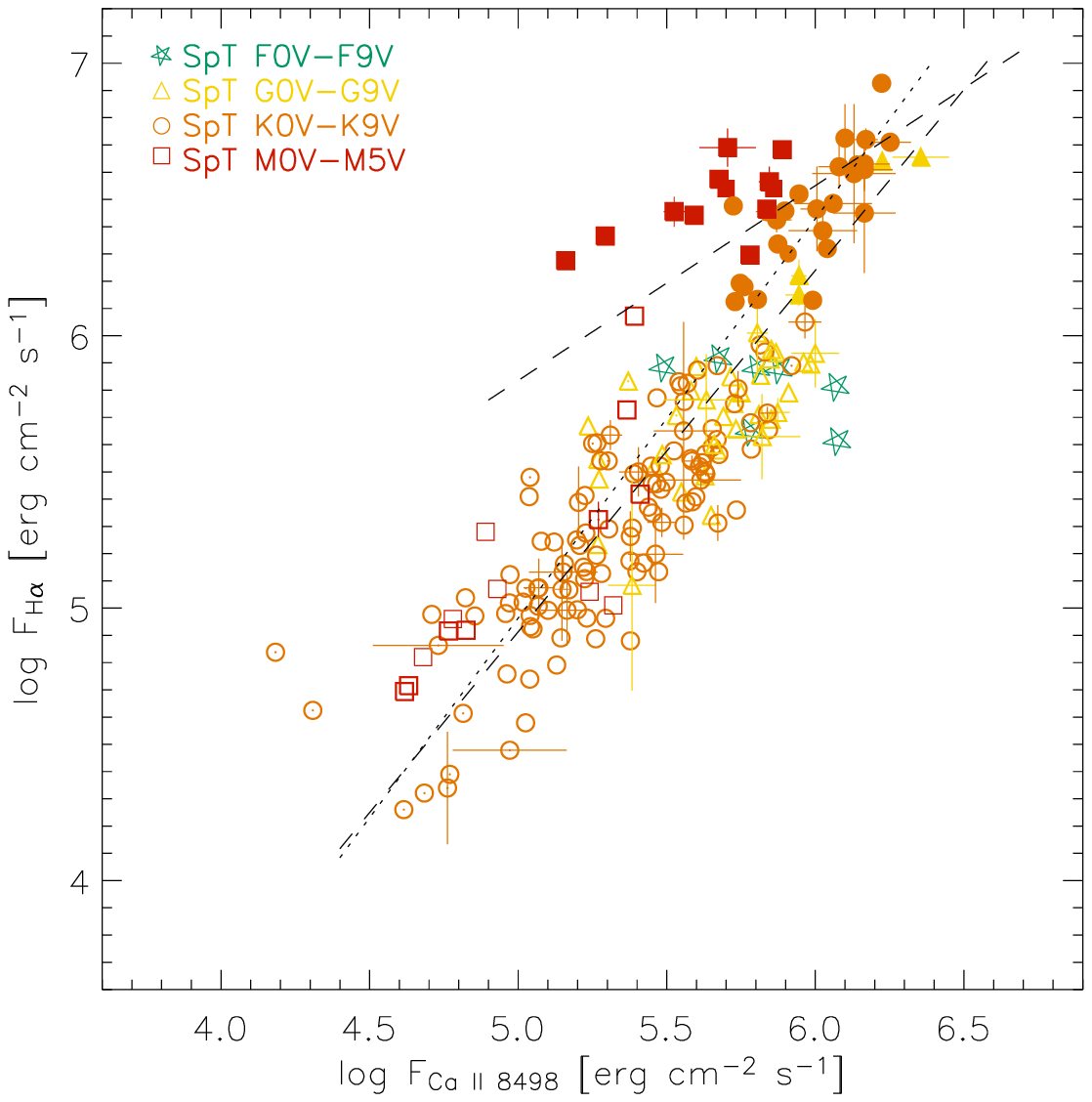}
\includegraphics[width=8cm, keepaspectratio]{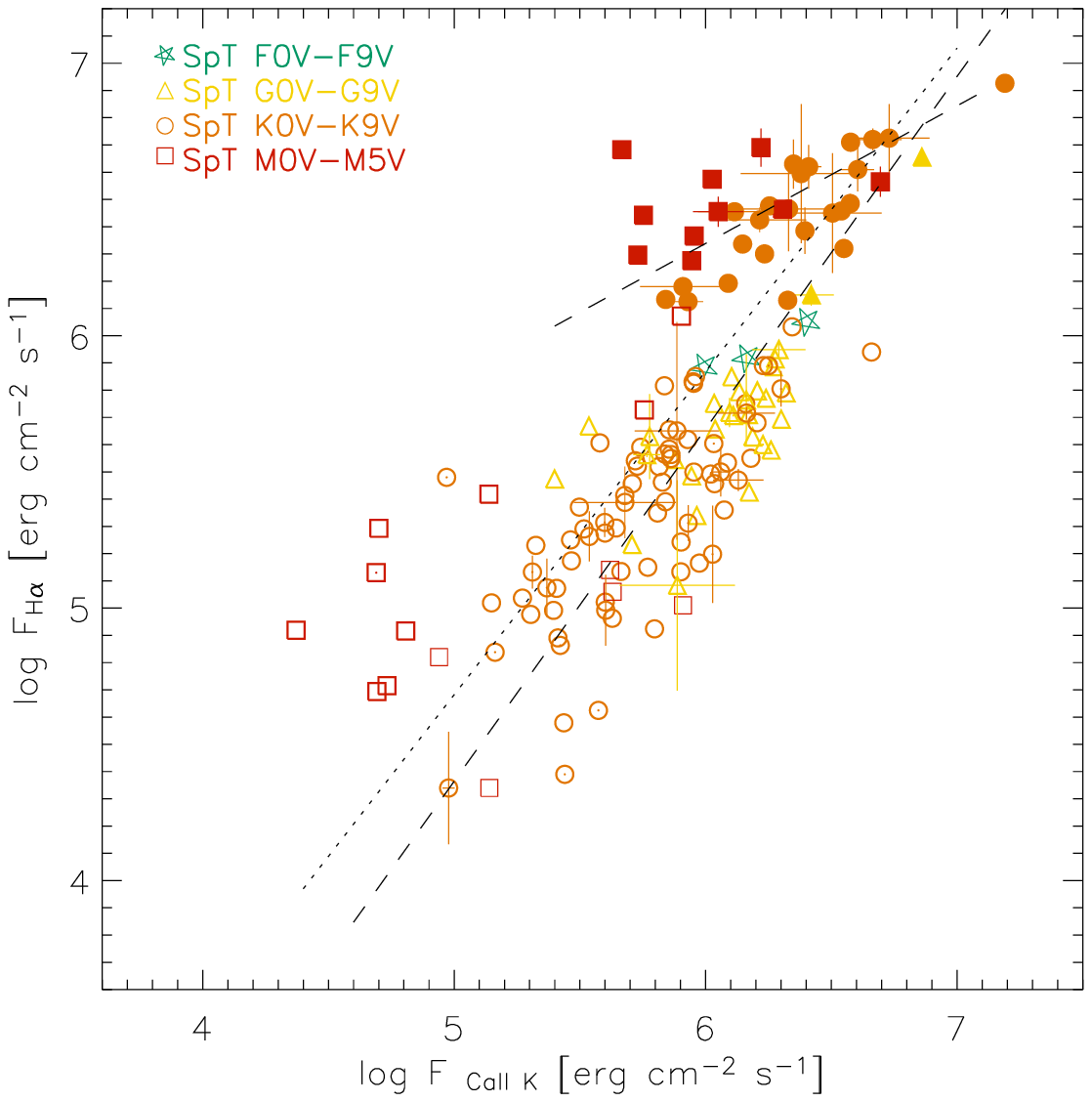}
\caption[]{Flux--flux relationships between H$\alpha$ and \ion{Ca}{ii} IRT $\lambda$8498 \AA\, ({\bf left})
and \ion{Ca}{ii} K ({\bf right}). Green stars represent F type stars, yellow triangles G type stars, orange circles K
type stars and red squares M type stars. Filled symbols are used for those stars which appear in the
upper branch H$\alpha$ vs. $B$--$V$ (those stars above the gap in Fig.~\ref{fig:gap}, and thus younger). 
The dotted line corresponds to the fit of all data whereas the dashed lines
represent the fit for each population of stars.}
\label{fig:flux_ha}
\end{figure*}
\begin{figure*}
\centering
\includegraphics[width=8cm, keepaspectratio]{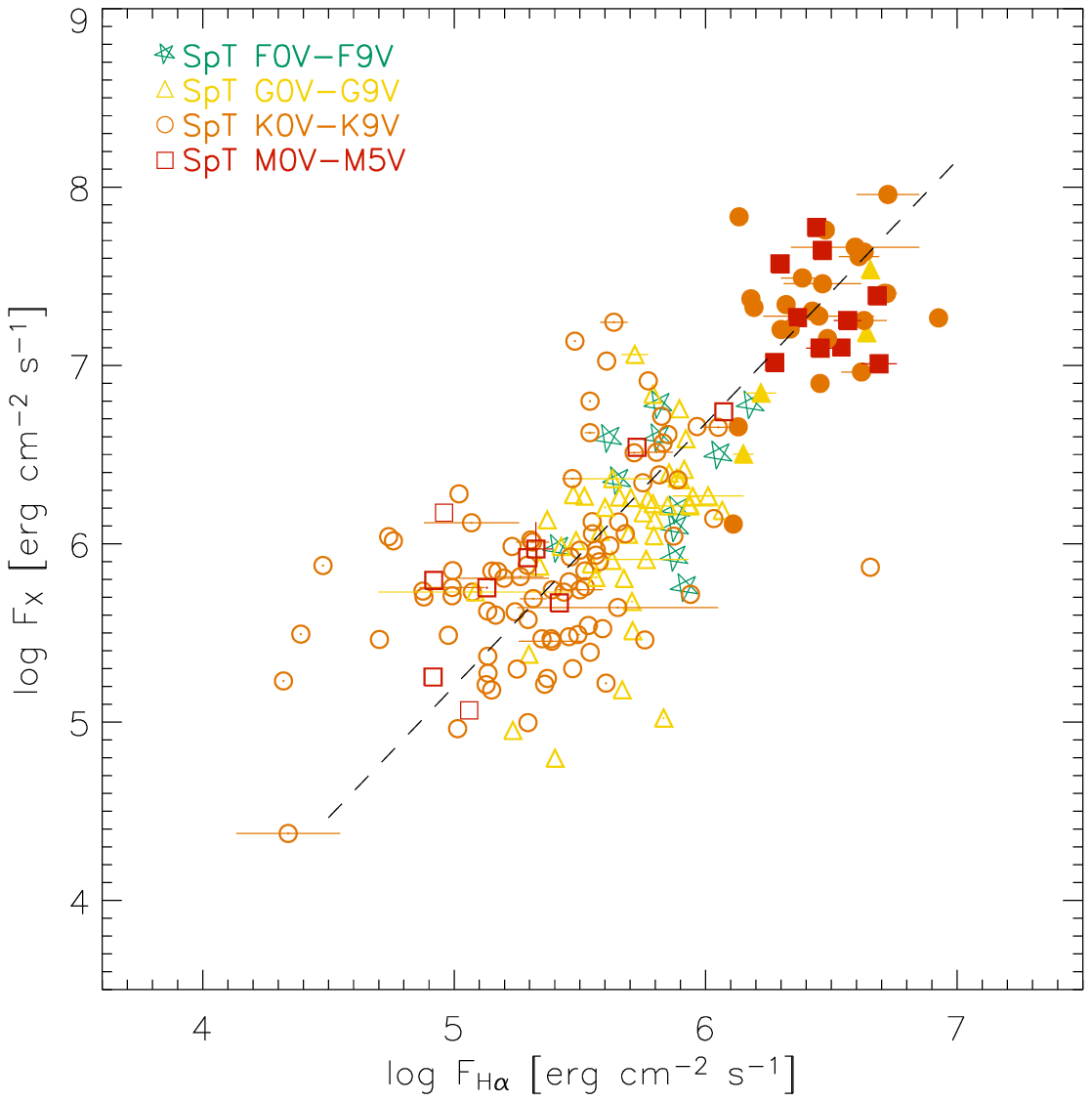}
\includegraphics[width=8cm, keepaspectratio]{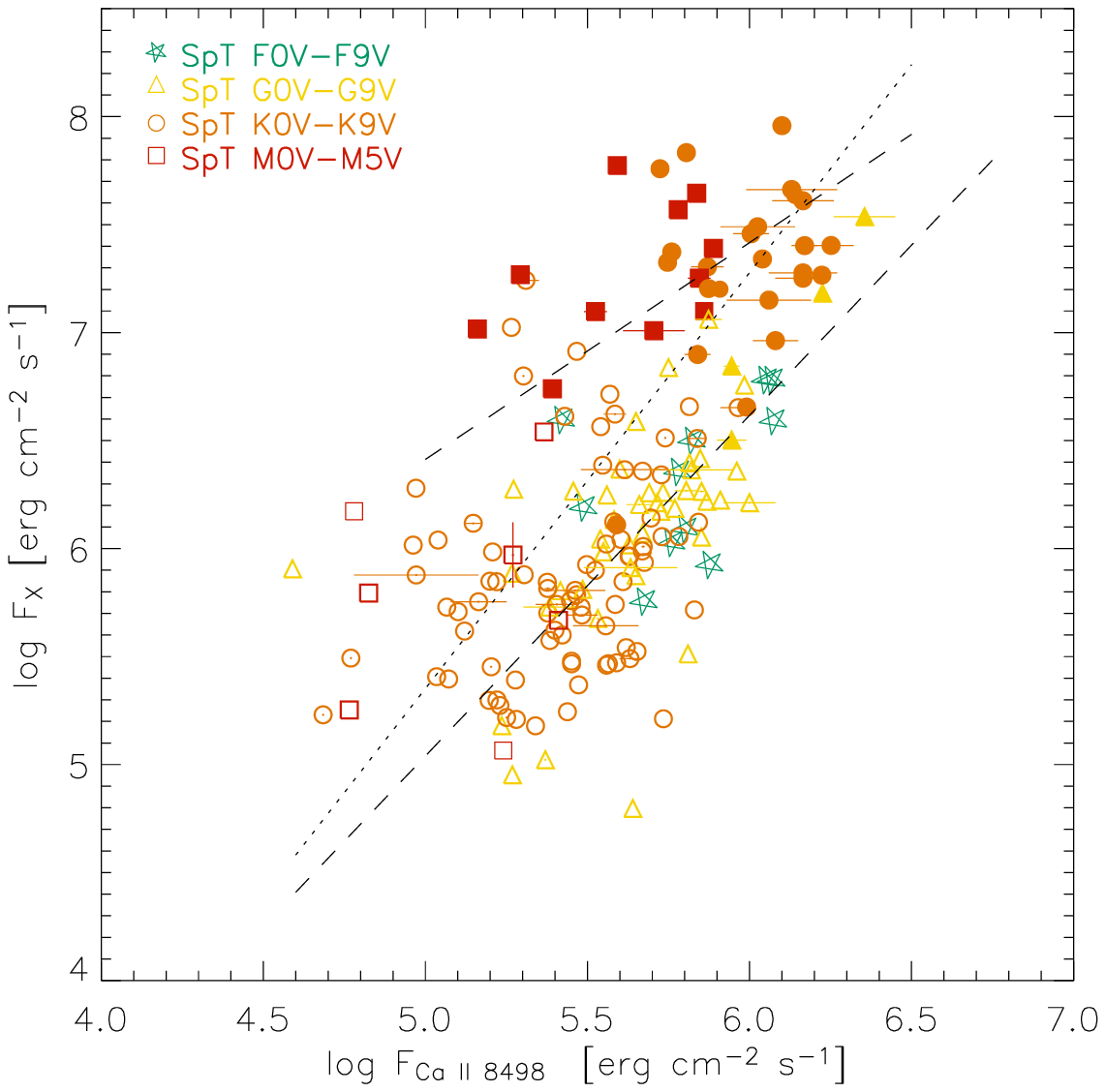}
\caption[]{Relationships between logarithmic X-ray surface fluxes and H$\alpha$ ({\bf left}), and 
\ion{Ca}{ii} $\lambda$8498 \AA\, ({\bf right}). Symbols, colours and line styles as in Fig.~\ref{fig:flux_ha}.}
\label{fig:flux_lx}
\end{figure*}

\section{Flux--flux relationships}
\label{sec_results}
In the following we address the subject of flux--flux relationships from a classical point of view and 
make no distinction between stars.
\subsection{Chromospheric flux--flux relationships}
\label{sec_flux--flux}

\begin{table*}
\caption{Linear fit coefficients for each flux--flux relationship found in this work and in previous ones.}
\label{tab:flux_rel}
\centering
\begin{tabular}{l l r  c c c}
\hline
\noalign{\smallskip}
 &  &  \multicolumn{2}{c}{This work} & & \multicolumn{1}{c}{Other studies}\\
\cline{3-4}   \cline{6-6}
\noalign{\smallskip}
$\log F_{\rm 1}$ & $\log F_{\rm 2}$ & \multicolumn{1}{c}{c$_{\rm 1}$} & c$_{\rm 2}$ && power law exponent\\
\noalign{\smallskip}
\hline
\noalign{\smallskip}
\multicolumn{2}{l}{\textbf{Chromosphere-Chromosphere}} \\
\noalign{\smallskip}
\hline
\noalign{\smallskip}
 \ion{Ca}{ii} K & \ion{Ca}{ii} H & 0.30 $\pm$ 0.17 & 0.97 $\pm$ 0.03 & & 1.00$^{\rm 1}$, 0.99$^{\rm 2}$\\
\noalign{\smallskip}
\ion{Ca}{ii} K & \ion{Ca}{ii} IRT {\tiny ($\lambda$8498 \AA)} & -0.29 $\pm$ 0.51 & 1.12 $\pm$ 0.09 & & 1.01$^{\rm 2}$\\
\noalign{\smallskip}
\ion{Ca}{ii} IRT ({\tiny $\lambda$8498 \AA}) & \ion{Ca}{ii} IRT ({\tiny $\lambda$8662 \AA}) & 
-0.13 $\pm$ 0.19 & 1.00$\pm$ 0.03 && ...\\
\noalign{\smallskip}
 \ion{Ca}{ii} IRT ({\tiny $\lambda$8498 \AA}) & \ion{Ca}{ii} IRT ({\tiny $\lambda$8542 \AA})& -0.33 $\pm$ 0.21
& 1.04 $\pm$ 0.03 & & 1.01$^{\rm 2}$\\
\noalign{\smallskip}
H$\alpha$  & \ion{Ca}{ii} IRT {\tiny ($\lambda$8498 \AA)} & -2.36 $\pm$ 0.35 & 1.46 $\pm$ 0.06 & & ...\\
\noalign{\smallskip}
 H$\alpha$ & \ion{Ca}{ii} K & -1.25 $\pm$ 0.52 & 1.19 $\pm$ 0.08 & & 1.13$^{\rm 1}$, 0.95$^{\rm 2}$,1.03$^{\rm 3}$,
1.12$^{\rm 4}$\\
\noalign{\smallskip}
\hline
\noalign{\smallskip}
\multicolumn{2}{l}{\textbf{Corona--Chromosphere}} \\
\noalign{\smallskip}
\hline
\noalign{\smallskip}
  X &  H$\alpha$ & -2.19 $\pm$ 0.41 & 1.48 $\pm$ 0.07 & & 2.11$^{\rm 4}$ \\ 
\noalign{\smallskip}
  X & \ion{Ca}{ii} IRT {\tiny ($\lambda$8498 \AA)} & -4.28 $\pm$ 0.77 & 1.98 $\pm$ 0.13 & & ...\\
\noalign{\smallskip}
\hline
\noalign{\smallskip}
\multicolumn{2}{l}{$^{\rm 1}$ L\'opez-Santiago et al. (2005)} & \multicolumn{2}{l}{$^{\rm 3}$ Montes et al. (1995b)}\\
\multicolumn{2}{l}{$^{\rm 2}$ Mart\'inez-Arn\'aiz et al. (2010)} & \multicolumn{2}{l}{$^{\rm 4}$ Montes et al. (1996a,b)}\\ 
\end{tabular}
\end{table*}

In Figs. \ref{fig:flux_all} and \ref{fig:flux_ha} we compare pairs of fluxes of 
different chromospheric lines for all the stars in our sample. 
For those stars observed more than once we plot the median value 
determined by us with all the observations and an error bar representing the maximum 
deviation from it.
Power-law relationships between pairs of lines
have been determined by fitting the data 
to a relation of the type

\begin{equation}
\label{eq:flux--flux}
\log F_{\rm 1} = c_{\rm 1} + c_{\rm 2} \log F_{\rm 2},
\end{equation}

where $F_{\rm 1}$ and $F_{\rm 2}$ are the fluxes of two different lines 
and $c_{\rm 1}$ and $c_{\rm 2}$ are the fitting parameters. 
We used the method explained in \citet{1990ApJ...364..104I}
to determine $c_{\rm 1}$, $c_{\rm 2}$ and their errors.
The bisector of the two possible ordinary least squares regressions (Y on X and 
X on Y) was determined in each case. The ordinary least squares bisector 
regression is the best solution when the goal is to estimate the underlying 
functional relation between the variables \citep[see][for a complete study of 
the problem]{1990ApJ...364..104I}.

In Table~\ref{tab:flux_rel} we present the results for $c_1$ and $c_2$.
This
table also lists the values of the slope ($c_2$) previously obtained in different 
studies by other authors \citep{1989ApJ...337..964S,1995A&A...294..165M,
1996A&A...312..221M,1996ASPC..109..657M,2005PhDT........14B,2010A&A...520A..79M}. 
Comparing columns \#5 and \#7 in Table~\ref{tab:flux_rel} we observe that  
the values obtained for the linear fit of the data in this work are compatible, taking
uncertainties into account, with those previously reported for single F, G and early-K stars 
for all chromospheric indicators. It is important to mention that while this study is based 
on single active stars, \citet{1995A&AS..114..287M} is focused on active binary stars. 
However, the differences obtained in the latter cases are not more noticeable than those 
obtained when one compares the results obtained in this work and other single star-based 
studies.

The values obtained 
for the slopes in the relationships between pairs of logarithmic fluxes follow the trend that has 
previously been reported, $i.e.$ 
the larger the difference in atmospheric height at which 
the compared lines form, the larger the value of the slope. This result has been obtained 
before using not only chromospheric indicators \citep{1996ASPC..109..657M}, but also transition region 
\citep{1986A&A...154..185O,1991A&A...251..183S,1996ASPC..109..657M} and coronal diagnostics 
\citep{1989ApJ...337..964S,1996ASPC..109..657M}.  

It is important to mention that after measuring excess $EW$, we applied the 
H$\alpha$ $EW$ criterion \citep{2003AJ....126.2997B} to ensure that no stars with an accretion disc were included. None of the stars 
in the sample had H$\alpha$ above the threshold and thus we can ensure that the measured $EW$ (and computed 
fluxes) are not affected by non-chromospheric H$\alpha$ emission.

\subsection{Chromospheric--coronal connection}
\label{sub:x-ray}

In addition to the flux--flux relationships between chromospheric 
activity diagnostics we have obtained empirical power laws between chromospheric and 
coronal (as given by X-rays) fluxes. Details on the search strategy and the 
conversion from count-rate to X-ray fluxes are given in Section \ref{sec:xray}. 
In Fig. \ref{fig:flux_lx} we present the empirical relationships between X-ray surface flux and  
H$\alpha$ (left panel) and  \ion{Ca}{ii} $\lambda8498$ \AA\,(right panel).
Table \ref{tab:flux_rel} includes the linear regression parameters for these 
relationships.

Due to the flux limit of the RASS, our sample of cross-matched X-ray sources is 
incomplete for low X-ray surface fluxes. For instance, at 20 pc, RASS is incomplete for 
$L_\mathrm{X} \le 10^{28}$ erg\,cm$^{-2}$\,s$^{-1}$ \citep[see][and Section 3.3]{2009A&A...499..129L}, 
which corresponds to $\log F_\mathrm{X} \sim 5.5$ (with flux in cgs) for a K dwarf. 
Note that some stars cross-correlated with an RASS counterpart show lower values of 
$\log F_\mathrm{X}$. However, we determined the relationships in the range 
$\log F_\mathrm{X} > 5.5$ to prevent any error in the determination of the relations due to this bias. 
In Fig.~\ref{fig:flux_lx}, the bias toward low $F_\mathrm{X}$ is clear, since only some stars 
with $\log F_\mathrm{X} \sim 5$ are present.

In general, the dispersion in these relationships is larger than that obtained  
between chromospheric lines. As mentioned in Sect.~\ref{sec:xray}, the uncertainties in the stellar 
radii estimations affect the X-ray surface flux determination and contributes to the dispersion observed 
in Fig. \ref{fig:flux_lx}. Another source of dispersion is the time variability of activity levels. 
While all the chromospheric indicators were obtained simultaneously, X-ray observations were performed at 
a different time. This result illustrates the importance of using simultaneous observations to build 
flux--flux relationships and avoid the time variability of activity levels.

Contrarily to what was observed for the chromospheric flux--flux relationships, our result with 
X-ray emission is quite different than that obtained by \citet{1996ASPC..109..657M}. In the latter study,  
the authors used only binary systems for their study, many of them with both stars emitting in X-rays
and/or in optical chromospheric lines. In some occasions, they could separate both components in 
their optical spectra, but never in X-rays (the authors used data from the \textit{ROSAT} not corrected
for binarity). This introduces systematical uncertainties and much more spread in the relationships. 
Therefore, the difference between our determined value for the slope of the coronal--chromospheric
flux relationships and that found by \citet{1996ASPC..109..657M} does not apply due to the 
systematic uncertainties introduced from determining X-ray emission from binary stars in the 
latter study. 

\section{The non universality of flux--flux relationships}
\subsection{Two distinct chromospheric emitter populations}
\label{sec:peculiarities_M}

\begin{table}
\caption{Linear fit coefficients for flux--flux relationships with the stars appearing in two different 
branches.}
\label{tab:flux_rel_2}
\scriptsize
\begin{tabular}{l l l r c c c}
\hline
\noalign{\smallskip}
$\log F_{\rm 1}$ & $\log F_{\rm 2}$ & Branch & \multicolumn{1}{c}{c$_{\rm 1}$} & c$_{\rm 2}$ \\
\hline
\noalign{\smallskip} 
H$\alpha$  & \ion{Ca}{ii} IRT {\tiny ($\lambda$8498 \AA)} & lower & -1.72 $\pm$ 0.35 & 1.33 $\pm$ 0.06 \\
\noalign{\smallskip}
& & upper & 2.26 $\pm$ 0.62 & 0.71 $\pm$ 0.10 \\
\noalign{\smallskip}
H$\alpha$ & \ion{Ca}{ii} K & lower & -2.10 $\pm$ 0.85 & 1.29 $\pm$ 0.14 \\
\noalign{\smallskip}
& & upper & 3.31 $\pm$ 0.43 & 0.50 $\pm$ 0.07\\
\noalign{\smallskip}
X & \ion{Ca}{ii} IRT {\tiny ($\lambda$8498 \AA)} & lower & -2.84 $\pm$ 0.84 & 1.58 $\pm$ 0.15\\
\noalign{\smallskip}
& & upper & 1.39 $\pm$ 0.42 & 1.00 $\pm$ 0.07\\
\noalign{\smallskip}
\hline
\end{tabular}
\end{table}

Despite having obtained values for the slope of the flux--flux relationships 
very similar to those found by other authors, a simple visual inspection of 
Fig.~\ref{fig:flux_ha} shows that 
some K-type (circles) and M-type (squares) stars clearly follow a different trend than 
that observed for other stars in some indicators. The linear relation still holds for these stars 
(dotted lines) but the slope is different from that obtained when all the stars are considered 
(see Table \ref{tab:flux_rel_2}). The deviation of these stars from 
the general relationships can be interpreted as an excess of H$\alpha$ 
emission with respect to \ion{Ca}{ii} K (right panel) or \ion{Ca}{ii} IRT (left panel).

Some authors have previously reported 
departures of dMe stars from general flux--flux relationships between chromospheric 
and transition-region indicators \citep[see][]{1986A&A...154..185O}, but their sample 
did not include enough stars to determine whether only dMe stars 
followed a different trend or it was an intrinsic behaviour of late-type stars. Different explanations for this 
behaviour have been proposed, including a different structure of the atmosphere of dMe stars 
\citep{1987A&A...177..143S} and a deficiency in the H \& K emission with re-radiation in other 
emission lines \citep{1989A&A...219..239R}, although the higher Balmer lines probably do not 
compensate the deficiencies observed in other chromospheric lines. Differences 
in the behaviour of flare dM stars when two chromospheric indicators are compared 
have been previously suggested but using a small sample of stars \citep{2005ESASP.560..775L}. 
This is the first time that a clear departure from the general chromospheric 
flux--flux relationships is observed for a sufficiently large sample of late-type stars.

It is important to mention here that, in contrast to most previous studies in which the chromospheric 
component of H$\alpha$ is taken to be simply the emission flux above a pseudo-continuum 
\citep{2003ApJ...583..451M,2004AJ....128..426W,2008AJ....135..785W}, in this 
work, H$\alpha$ EWs were measured in the subtracted spectrum. This fact ensures that the measured 
fluxes are only chromospheric fluxes, giving full credibility to the deviations observed in the 
flux--flux relationships.

The peculiar behaviour of these stars extends when we compare 
X-ray fluxes to those in chromospheric calcium lines (see Fig. \ref{fig:flux_lx}). 
Given that X-ray and H$\alpha$ fluxes present an almost linear relation (see Table 
\ref{tab:flux_rel}) it is not surprising that those stars which deviate from the main 
flux--flux relationships when H$\alpha$ is one of the diagnostics, also  
deviate when X-ray emission is considered. Note that \citet{1989A&A...219..239R} found 
that dMe stars deviated from the $F_{\rm X}$-$F_{\rm H\alpha}$ relationship. However, in the latter 
work, they subtracted the contribution of the photosphere in F and G stars but not in the M stars.

For the three diagrams in which two branches are clearly observed, we 
have determined flux--flux relationships for each branch separately. The results are
shown in Table~\ref{tab:flux_rel_2}. Notice that M stars in Fig. \ref{fig:flux_ha} 
are situated even above the obtained linear fit of the upper branch stars. These stars 
might follow a slightly different trend. 
It is important to note that the new relationships for the lower 
branches show different slopes than those obtained when all the stars 
are used for the fitting, especially for the most active stars. This difference 
should be taken into account when working with samples of late-type active stars.

\subsubsection{The nature of the two populations}

\begin{figure*}
\centering
\includegraphics[width=8cm, keepaspectratio]{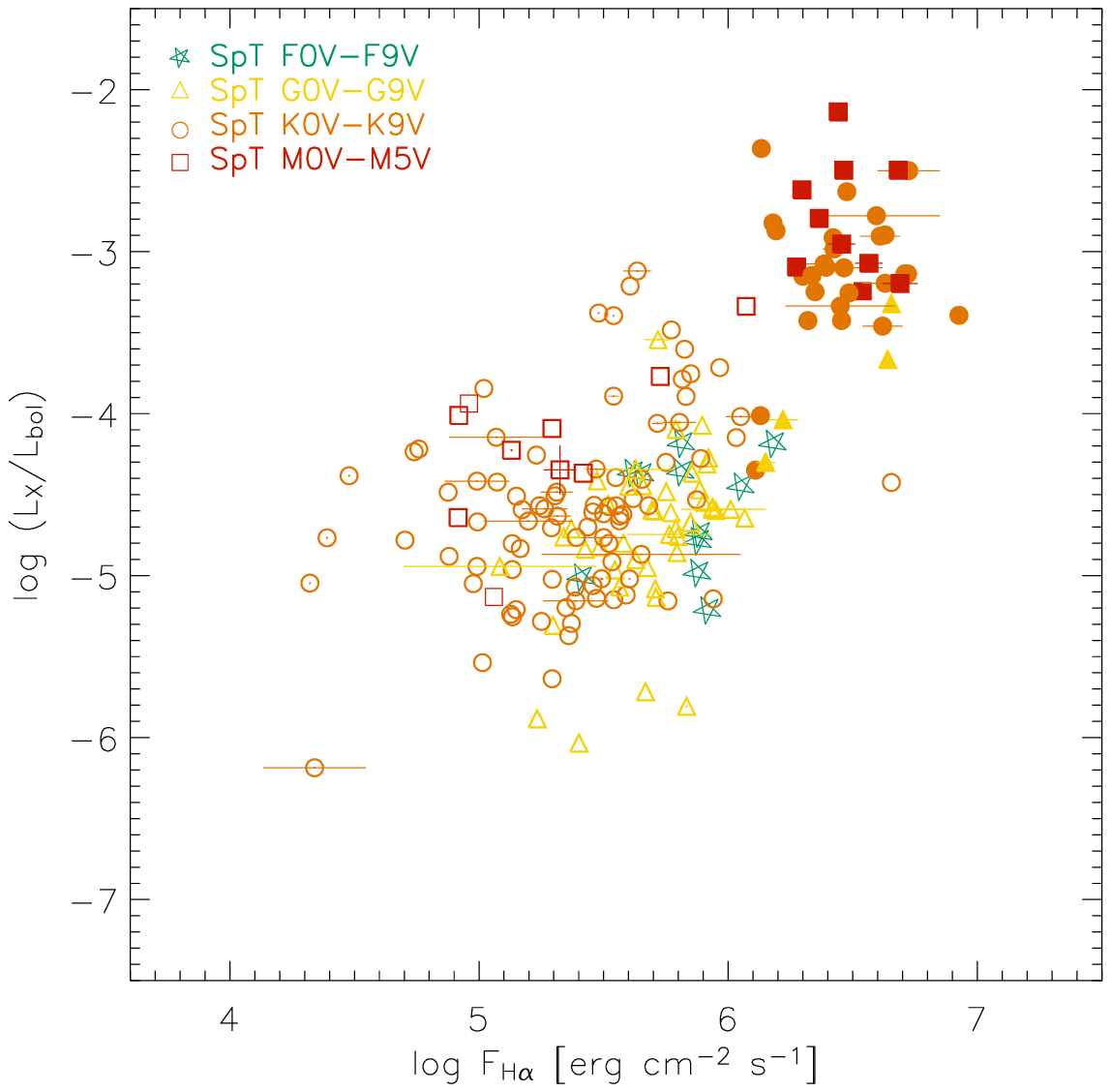}
\includegraphics[width=8cm, keepaspectratio]{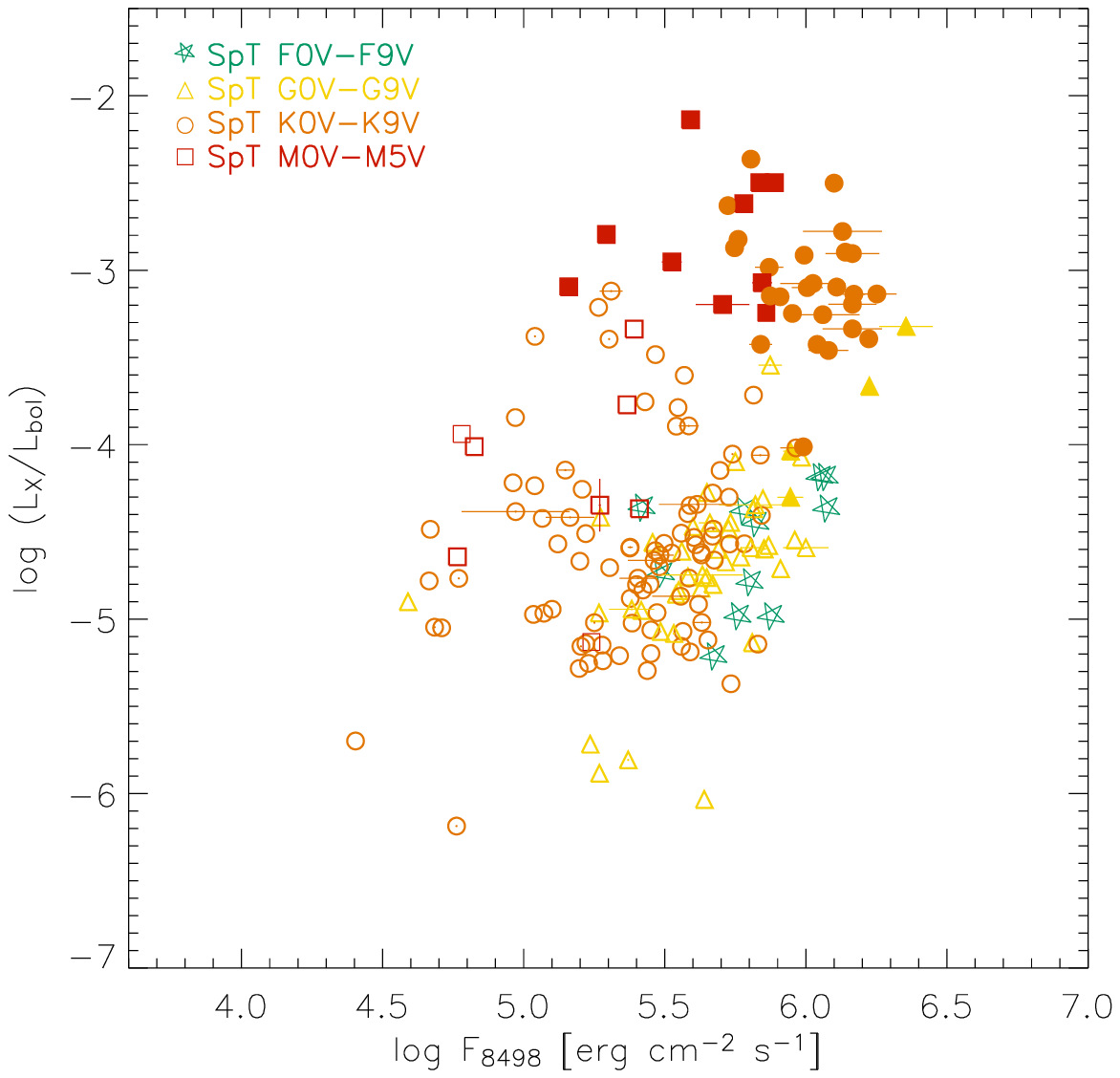}
\caption[]{X-ray luminosity vs. H$\alpha$ excess emission ({\bf left}) and 
\ion{Ca}{ii} IRT $\lambda$8498 \AA\,  ({\bf right}). Symbols and colours as in Fig.~\ref{fig:flux_ha}.}
\label{fig:lx}
\end{figure*}

We have investigated the nature of the apparent two populations of chromospheric 
emitters. In Fig. \ref{fig:lx}, we plot $\log L_\mathrm{X}/L_\mathrm{bol}$ versus
$\log F_\mathrm{H\alpha}$ and $\log F_{8498}$. The $L_\mathrm{X}/L_\mathrm{bol}$ ratio 
is a good estimator of magnetic activity in late-type stars. It reaches a maximum 
$\sim 10^{-2,-3}$ for (very active) high X-ray emitters and decreases to $10^{-5,-6}$ 
for less magnetically active stars. This behaviour is related to stellar rotation. It has been 
observed that X-ray emission in fast rotators is saturated \citep[see][and references therein]{2003A&A...397..147P}.
This saturation has been 
proved to be independent of stellar mass and rotation rate 
\citep{1995A&A...294..515H,2000MNRAS.318.1217J}. 
The physical phenomena behind this saturated regime is still controversial, but the 
transition boundary between both regimes has been found to be dependent on rotational period 
for a given spectral type \citep{2003A&A...397..147P}, and thus dependent on age due to 
the time-dependence of stellar rotation \citep{2003ApJ...586..464B,2008ApJ...687.1264M}.
Thus, $L_\mathrm{X}/L_\mathrm{bol}$ can be used as a measure
of the level of stellar magnetic activity and an estimator of age. 
Fig.~\ref{fig:lx} shows that stars with higher 
chromospheric H$\alpha$ surface fluxes are in the X-ray saturation regime. 
Those stars are mainly situated in the upper branch in Fig.~\ref{fig:flux_ha}. 
Therefore, the stars conforming the upper branch in the H$\alpha$-\ion{Ca}{ii} 
flux--flux relationships are mainly saturated X-ray emitters. This hypothesis was 
suggested by \citet{2009A&A...499..129L} in a study of a sample 
of young stars which were possible members of young stellar kinematic groups. 
In that work, the authors observed that there were two well-defined branches of 
H$\alpha$ emitters in the $\log F_\mathrm{H\alpha}$--($V-I$) diagram.

To confirm that these stars have rotation rates in agreement with having a saturated 
X-ray emission we have compiled the photometric periods when available. 
We list the values in Table \ref{tab:upper_branch}. With the exception of HIP 23309, DK Leo and HD 135363, all the upper 
branch stars for which photometric periods were measured, have P$_{\rm phot} < 4.5 $ days. This 
corresponds to the saturated regime for each mass range according to the results 
obtained by \citet{2003A&A...397..147P}.

\begin{figure*}
\centering
\includegraphics[width=8cm, keepaspectratio]{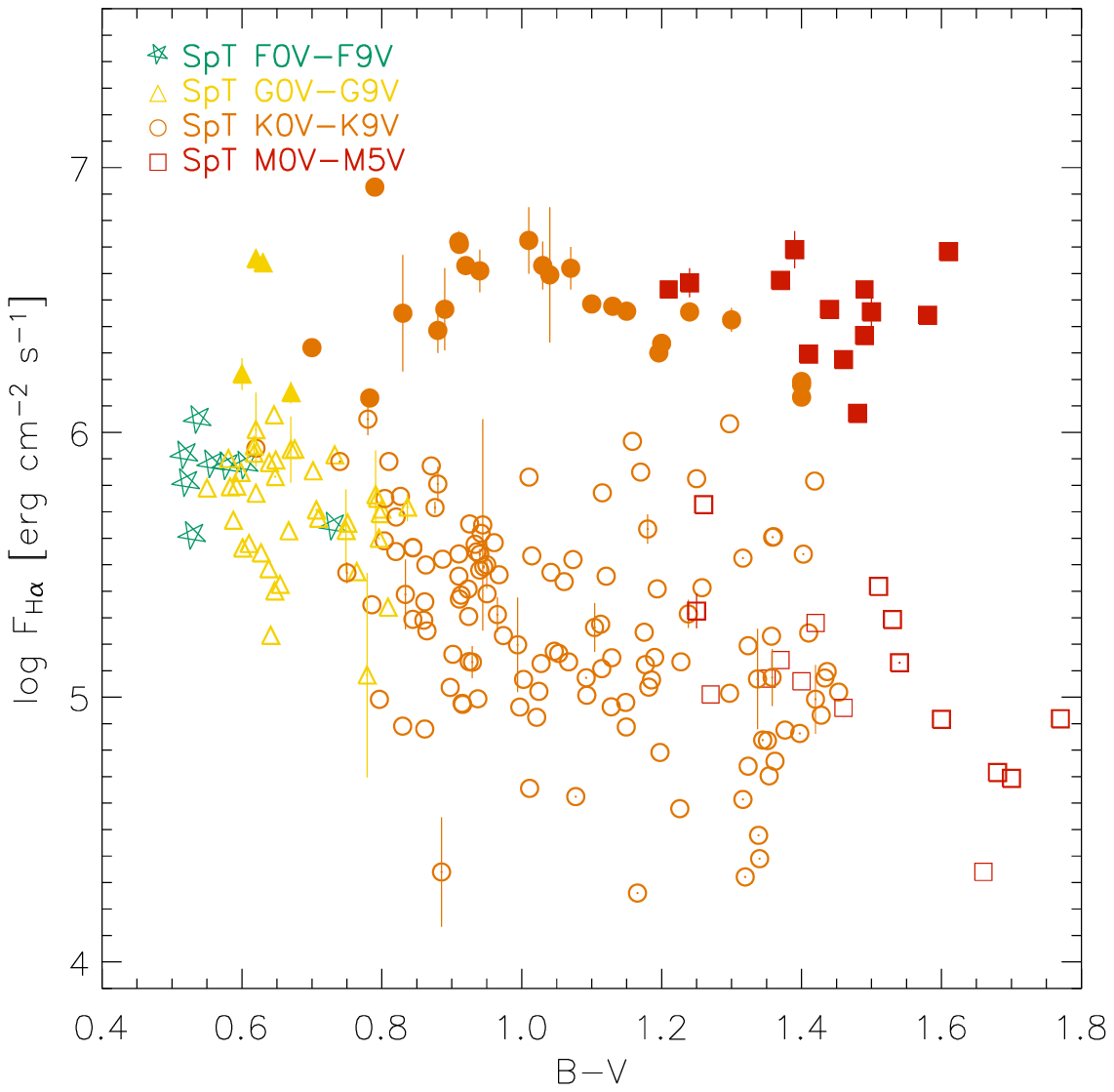}
\includegraphics[width=8cm, keepaspectratio]{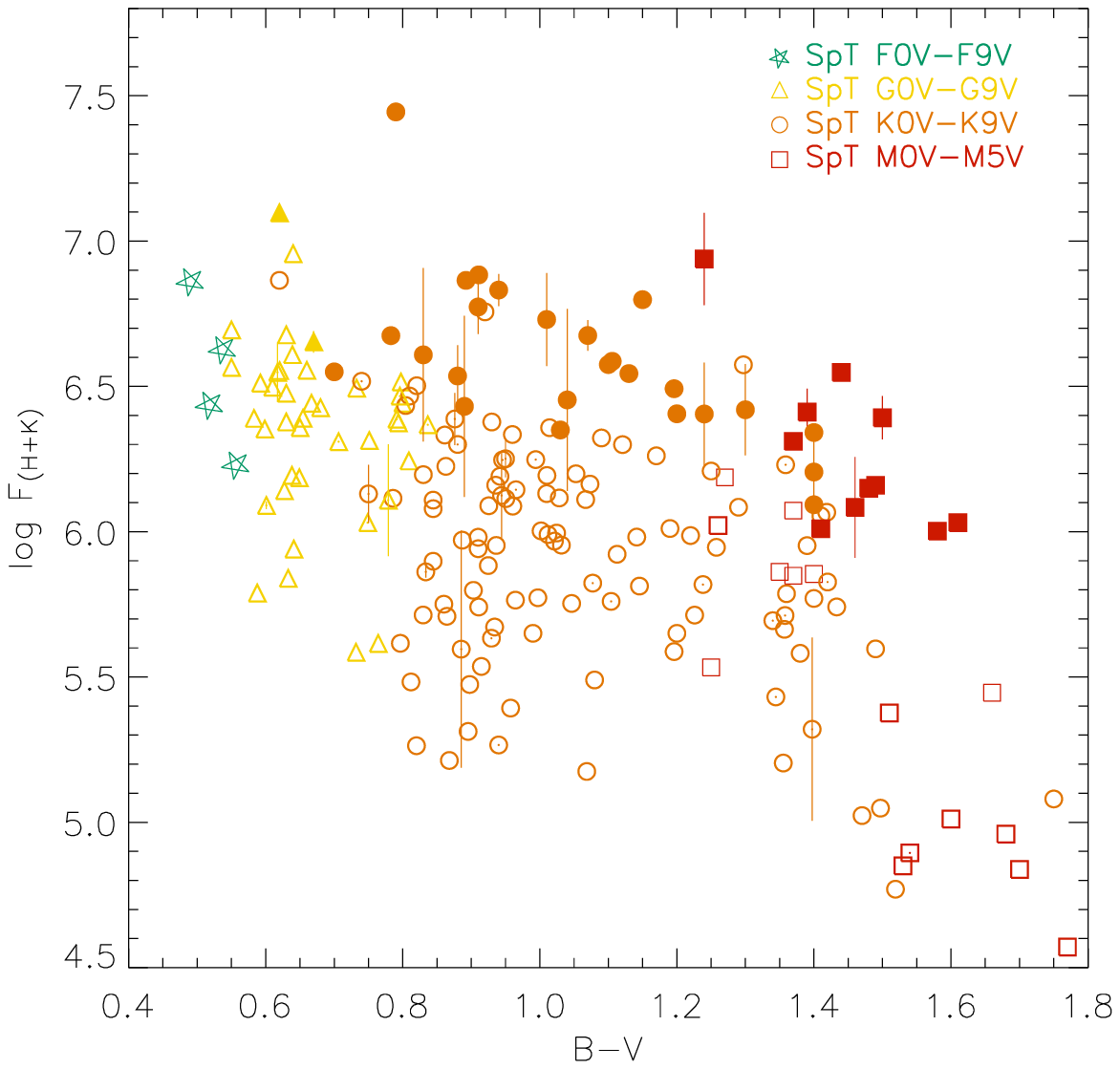}
\caption[]{H$\alpha$ excess emission vs. colour index $B$--$V$ ({\bf left}). Total excess emission 
in  \ion{Ca}{ii} H \& K vs. colour index $B$--$V$ ({\bf right}). Filled symbols are used for those 
stars that conform the upper region of the gap. Colors as in Fig.~\ref{fig:flux_ha}.}
\label{fig:gap}
\end{figure*}

Given that the two-population distribution resembles the segregation observed by 
\citet{1980PASP...92..385V} in \ion{Ca}{ii} H \& K, we have studied whether there is a relation 
between the discrepancy in H$\alpha$ emission and the position of these stars in a 
flux--colour diagram as that presented by Vaughan \& Preston.
We have studied the position of the upper branch 
stars in an activity--colour index diagram (see left  panel in Fig. \ref{fig:gap}). 
We have plotted the emission flux in H$\alpha$ (left) and the total flux in 
\ion{Ca}{ii} (right) H \& K lines vs. $B$--$V$. There is a clear gap between stars with 
high H$\alpha$ emission and those with lower activity levels, 
particularly for K and M stars. 
We have used filled 
symbols  to mark these stars in all the plots. We note that those stars in 
the region on top of the gap in the H$\alpha$-($B$--$V$) relation are the ones that deviate from 
the general flux--flux relationship between H$\alpha$ and the calcium lines. Note also 
that the gap is not as noticeable when the used activity index is the total flux in \ion{Ca}{ii} 
H \& K instead of H$\alpha$, yet those stars with abnormal H$\alpha$ emission are in the upper 
region in this plot, too. A similar result was found by LS10.
In Table~\ref{tab:upper_branch}, we list the stars that belong to the upper branch
in Fig.~\ref{fig:gap}. For them, we give values of $\log F_\mathrm{H\alpha}$ and
$\log F_\mathrm{X}$.

\begin{table*}
\caption{Stellar parameters of the stars in the `upper branch'.}
\label{tab:upper_branch}
\scriptsize
\centering
\begin{tabular}{ l l l l l l c c l c}
\hline
\noalign{\smallskip}
{Name} &
{SpT} &
\multicolumn{1}{c}{$B$--$V$} & \multicolumn{1}{c}{RA} & \multicolumn{1}{c}{DEC} &
{Source} & \multicolumn{1}{c}{$\log{F_{\rm H\alpha}}$} &
{$\log{F_{\rm X}}$}  & Notes on the stellar age (1) & P$_{\rm phot}$ (2)\\
\noalign{\smallskip}
 &
&  & \multicolumn{1}{c}{(hh mm ss)} & \multicolumn{1}{c}{($\deg$ ' '')} &
& \multicolumn{1}{c}{(erg cm$^{\rm -2}$ s$^{\rm -1}$)} & \multicolumn{1}{c}{(erg cm$^{\rm -2}$ s$^{\rm -1}$)} & 
\multicolumn{1}{c}{} & (days)\\
\hline
\noalign{\smallskip}
PW And	&	K2V	&	1.04	&	00 18 20.89	&	+30 57 22.2	&	LS10	&	6.54	$\pm$	0.06	&	7.66	&	AB Dor$^{\rm 1}$, flare star	&	1.74$^{\rm E}$	\\
         QT And	&	K2V	&	0.92	&	00 41 17.34	&	+34 25 16.9	&	LS10	&	6.63	$\pm$	0.07	&	7.63	&	Ca, EW (LiI) =132.2 m\AA\, $^{\rm 2, b}$	&	2.86$^{\rm M}$	\\
      BD+17 232	&	K4V	&	1.01	&	01 37 39.42	&	+18 35 32.9	&	LS10	&	6.70	$\pm$	0.11	&	7.96	&	LA, EW (LiI) =  405.6 m\AA\,$^{\rm 2, a}$, flare star	&	...	\\
V577 Per	&	K1V	&	0.70	&	03 33 13.49	&	+46 15 26.5	&	LS10	&	6.32	$\pm$	0.12	&	7.34	&	AB Dor$^{\rm 1}$	&	1.45$^{\rm J}$	\\
HD 25457	&	F6V	&	0.52	&	04 02 36.74	&	-00 16 08.1	&	LS10	&	6.18	$\pm$	0.19	&	6.78	&	AB Dor$^{\rm 1}$	&	...	\\
V834 Tau	&	K3V	&	1.10	&	04 41 18.85	&	+20 54 05.4	&	LS10	&	6.48	$\pm$	0.17	&	7.15	&	Uma, EW (LiI) = 56.5 m\AA\, $^{\rm 2, b}$, flare star	&	3.94$^{\rm J}$	\\
V1005 Ori	&	M1.5V	&	1.41	&	04 59 34.83 	&	+01 47 00.6	&	FEROS05	&	6.30	$\pm$	0.02	&	7.57	&	$\beta$ Pic$^{\rm 5}$, flare star	&	4.4$^{\rm L,O}$	\\
HIP 23309	&	M0.5V	&	1.40	&	05 00 47.13	&	-57 15 25.4	&	FEROS05	&	6.19	$\pm$	0.02	&	7.32	&	$\beta$ Pic$^{\rm 5}$	&	8.6$^{\rm L}$	\\
V371 Ori	&	M2.5V	&	1.58	&	05 33 44.79	&	+01 56 43.4 	&	FEROS05	&	6.44	$\pm$	0.01	&	7.77	&	YD$^{\rm 4 }$,flare star	&	...	\\
UY Pic B	&	K5V	&	0.79	&	05 36 55.07	&	-47 57 47.9	&	FEROS05	&	6.93	$\pm$	0.04	&	7.27	&	AB Dor$^{\rm 1}$	&	4.52$^{\rm C}$	\\
AO Men	&	K4IV	&	1.13	&	06 18 28.21	&	-72 02 41.5	&	FEROS05	&	6.48	$\pm$	0.01	&	7.76	&	$\beta$ Pic$^{\rm 5}$	&	2.67$^{\rm O}$	\\
     BD+20 1790	&	K5V	&	1.07	&	07 23 43.59	&	+20 24 58.7	&	LS10	&	6.62	$\pm$	0.07	&	6.96	&	AB Dor$^{\rm 1}$, flare star	&	2.8$^{\rm D,M}$	\\
V372 Pup	&	M1.5IV	&	1.44	&	07 28 51.37	&	-30 14 48.5	&	FEROS05	&	6.46	$\pm$	0.01	&	7.64	&	AB Dor$^{\rm 1}$	&	1.64$^{\rm G}$	\\
BD+21 2462	&	K3V	&	1.01	&	07 43 48.49	&	-36 13 06.5	&	LS10	&	6.11	$\pm$	0.23	&	6.11	&	IC, EW (LiI) = 40.4 m\AA\, $^{\rm 2, b}$	&	...	\\
YZ CMi	&	M4.5V	&	1.61	&	07 44 40.17	&	+03 33 08.8	&	FEROS05	&	6.68	$\pm$	0.04	&	7.39	&	LA$^{\rm 4}$, flare star	&	2.78$^{\rm P}$	\\
FP Cnc	&	K7V	&	1.40	&	08 08 56.40	&	+32 49 11.2	&	LS10	&	6.18	$\pm$	0.06	&	7.37	&	LA$^{\rm 2}$, flare star	&	...	\\
HD 77407	&	G0V	&	0.60	&	09 03 27.08	&	+37 50 27.5	&	LS10	&	6.22	$\pm$	0.20	&	6.84	&	LA, EW (LiI) = 161.3 m\AA\, $^{\rm 2, b}$	&	...	\\
LQ Hya	&	K0V	&	0.91	&	09 32 25.57	&	-11 11 04.7	&	MA10/LS10	&	6.71	$\pm$	0.02	&	7.40	&	YD, EW (LiI) =  240.3 m\AA\,$^{\rm 2,a}$	&	1.60$^{\rm H,Q}$	\\
DX Leo	&	K0V	&	0.78	&	09 32 43.76	&	+26 59 18.7	&	MA10/LS10	&	6.13	$\pm$	0.03	&	6.65	&	LA, EW (LiI) = 192.4$^{\rm 3, b}$, 176.3$^{\rm 2, b}$ m\AA\,	&	6 $^{\rm A}$	\\
      DK Leo	&	K7V	&	1.24	&	10 14 19.18	&	+21 04 29.6	&	LS10	&	6.46	$\pm$	0.08	&	6.90	&	LA$^{\rm 2}$, flare star	&	7.98$^{\rm N}$	\\
         AD Leo	&	M3.5V	&	1.50	&	10 19 36.28	&	+19 52 12.1	&	LS10	&	6.45	$\pm$	0.08	&	7.10	&	Cas$^{\rm 2}$, flare star	&	2.60$^{\rm N}$	\\
V857 Cen	&	M4.5V	&	1.54	&	11 31 46.51	&	-41 02 47.2	&	FEROS05	&	6.11	$\pm$	0.02	&	7.97	&	HS$^{\rm 2}$, flare star	&	...	\\
EQ Vir	&	K4V	&	1.20	&	13 34 43.21	&	-08 20 31.3	&	FEROS05/MA10	&	6.34	$\pm$	0.01	&	7.20	&	IC$^{\rm 2}$, flare star	&	3.96$^{\rm  B}$	\\
EK Dra	&	G1.5V	&	0.63	&	14 39 00.21	&	+64 17 29.9	&	LS10	&	6.46	$\pm$	0.04	&	7.18	&	LA, subgroup B4 $^{\rm 1}$	&	2.79$^{\rm J}$	\\
IU Dra	&	G0V	&	0.67	&	15 05 49.90	&	+64 02 49.9	&	LS10	&	6.15	$\pm$	0.29	&	6.50	&	HS, EW (LiI) = 144.9 m\AA\, $^{\rm 2, b}$	&	4.45$^{\rm J}$	\\
HD 135363	&	K0V	&	0.94	&	15 07 56.26	&	+76 12 02.7	&	LS10	&	6.65	$\pm$	0.07	&	7.61	&	HS, EW (LiI) = 198.2 m\AA\, $^{\rm 2, b}$	&	7.24$^{\rm G}$	\\
HD 139751	&	K3/K4V	&	1.40	&	15 40 28.39	&	-18 41 46.2	&	FEROS05	&	6.13	$\pm$	0.02	&	7.83	&	AB Dor$^{\rm 1}$	&	3.7$^{\rm L}$	\\
CR Dra	&	M1V	&	1.24	&	16 17 05.39	&	+55 16 09.1 	&	LS10	&	6.55	$\pm$	0.08	&	7.25	&	YD$^{\rm 2}$, flare star	&	...	\\
V1054 Oph	&	M4V	&	1.49	&	16 55 28.75	&	-08 20 10.8	&	FEROS05	&	6.36	$\pm$	0.01	&	7.26	&	HS$^{\rm 2}$, flare star	&	...	\\
V647 Her	&	M3.5V	&	1.46	&	17 19 54.20	&	+26 30 03.0	&	LS10	&	6.28	$\pm$	0.05	&	7.02	&	HS$^{\rm 2}$, flare star	&	1.34$^{\rm J}$	\\
V889 Her	&	G2V	&	0.62	&	18 34 20.10	&	18 41 24.2	&	LS10	&	6.65	$\pm$	0.15	&	7.54	&	LA, EW (LiI) = 208.4 m\AA\, $^{\rm 2, a}$	&	1.34$^{\rm F}$	\\
  2RE J1846+191	&	K4V	&	1.49	&	18 46 09.34	&	+19 12 14.9	&	LS10	&	6.12	$\pm$	0.07	&	...	&	YD$^{\rm 2}$	&	...	\\
LO Peg	&	K3V	&	1.03	&	21 31 01.71	&	+23 20 07.4	&	LS10	&	6.65	$\pm$	0.08	&	7.25	&	AB Dor$^{\rm 1}$	&	0.42$^{\rm I,L}$	\\
       V383 Lac	&	K1V	&	0.83	&	22 20 07.03	&	+49 30 11.8	&	LS10	&	6.44	$\pm$	0.12	&	7.28	&	LA, EW (LiI) =  260.0 m\AA\,$^{\rm 2, a}$	&	2.42$^{\rm J}$	\\
        GJ 856 B	&	M1V	&	1.49	&	22 23 30.00	&	+32 27 00.0	&	LS10	&	6.54	$\pm$	0.06	&	...	&	LA, subgroup B4 $^{\rm 1}$	&	...	\\
     BD+17 4799	&	K0V/IV	&	0.88	&	22 44 41.54	&	+17 54 18.3	&	LS10	&	6.37	$\pm$	0.12	&	7.49	&	LA, EW (LiI) = 247.6 $^{\rm 2,a}$	&	...	\\
EV Lac	&	M3.5V	&	1.39	&	22 46 49.73	&	+44 20 02.4	&	LS10	&	6.69	$\pm$	0.04	&	7.01	&	LA$^{\rm 2}$, flare star	&	4.38$^{\rm N}$	\\
GJ 9809	&	M0V	&	1.21	&	23 06 04.84	&	+63 55 34.4	&	LS10	&	6.54	$\pm$	0.05	&	7.10	&	AB Dor$^{\rm 1}$, flare star	&	4.50$^{\rm G}$	\\
V368 Cep	&	K0V	&	0.89	&	23 19 26.63	&	+79 00 12.7	&	LS10	&	6.44	$\pm$	0.10	&	7.46	&	LA, EW (LiI) = 205.4 m\AA\, $^{\rm 2, b}$	&	2.74$^{\rm H}$	\\
\hline
\noalign{\smallskip}
\multicolumn{10}{l}{(1) $^{\rm 1}$\citet{2006ApJ...643.1160L}, $^{\rm 2}$LS10, $^{\rm 3}$\citet{2010A&A...521A..12M}, $^{\rm 3}$\citet{2001MNRAS.328...45M}
$^{\rm 5}$\citet{2008hsf2.book..757T}. $^{\rm a}$ Consistent with being younger than the Pleiades,}\\
\multicolumn{10}{l}{$^{\rm b}$ Consistent with the age of the Pleiades. Ca: Castor, HS: Hyades SC, IC: IC2391, LA: Local Association, UMa: Ursa Major MG, 
YD: Young Disk.}\\
\noalign{\smallskip}
\multicolumn{10}{l}{(2) $^{\rm A}$\citet{1996ApJ...457L..99B}, $^{\rm B}$\citet{1977AJ.....82..490B}, $^{\rm C}$\citet{1999A&AS..138...87C}, 
$^{\rm D}$\citet{2010A&A...512A..45H}, $^{\rm E}$\citet{1990ApJS...74..225H}, $^{\rm F}$\citet{2008A&A...488.1047J}, $^{\rm G}$\citet{2002MNRAS.331...45K},}\\
\multicolumn{10}{l}{$^{\rm H}$\citet{2004A&A...417.1047K}, $^{\rm I}$\citet{1999MNRAS.307..685L}, $^{\rm J}$\citet{2001A&A...366..215M}, 
$^{\rm K}$\citet{2003A&A...410..671M}, $^{\rm L}$\citet{2010A&A...520A..15M}, $^{\rm M}$\citet{2007A&A...467..785N}, $^{\rm N}$\citet{2003A&A...397..147P},}\\
\multicolumn{10}{l}{$^{\rm O}$\citet{2003AcA....53..341P},$^{\rm P}$\citet{2001ASPC..223..292S}, $^{\rm Q}$\citet{1997A&AS..125...11S}.}
\end{tabular}
\end{table*}
\normalsize

\subsubsection{Ages and dynamo}

Diagrams such as the ones presented in Fig.~\ref{fig:gap}
have been interpreted as the
result of the rapid decay of the magnetic flux with age rather than to a discontinuity in 
emission levels. The saturation in the chromospheric emission of young stars 
\citep{1984ApJ...276..254H, 1987ARA&A..25..271H} would result in the 
concentration of stars above the Vaughan--Preston 
gap. Therefore, the gap would separate young stars from older ones 
\citep{1984ApJ...279..763N,2004ARA&A..42..685Z,2007ApJ...657..486B,2009A&A...499L...9P}. 

Several authors have suggested that the gap actually separates stars with two different 
dynamo regimes and that this dynamo could change over the star's lifetime. In a recent 
study, \citet{2007ApJ...657..486B} concludes that there are two different dynamos at work in 
the two branches observed in the \textit{cycle period} versus \textit{rotation period} diagram. 
The existence of two dynamos was previously 
proposed by \citet{1981PASP...93..537D} as an explanation for the Vaughan--Preston gap. 
They also suggested that this could produce a change in the morphology 
of the magnetic field between early-type  and late-type stars.
Although this interpretation is still controversial, it  could be a plausible explanation for 
the  two clearly independent populations of stars we find when the H$\alpha$ 
emission is considered.

To investigate if the upper branch is formed by young stars, we studied age 
indicators in our sample. In particular, we used data from \citet{2006ApJ...643.1160L}, LS10,  
\citet{2010A&A...521A..12M}, \citet{2001MNRAS.328...45M}, and \citet{2008hsf2.book..757T} 
to determine if the star was a member of a young stellar association or moving group. 
In Table~\ref{tab:upper_branch} we give details on the membership of each star in the 
upper branch of Fig.~\ref{fig:gap} in any young group.
>From the 39 stars belonging to the upper branch, nine are members of the AB Dor moving group
\citep[$\sim 50$ Myr;][]{2004ARA&A..42..685Z}, three are members of the $\beta$ Pic association 
\citep[$\sim$ 12-20 Myr;][]{1999ApJ...520L.123B,2004ARA&A..42..685Z}, and two stars are members 
of the subgroup B4 of the Local Association \citep[$\sim$ 80-120 Myr;][]{2006ApJ...643.1160L}.
 
For those stars in LS10 and MA10, the preliminary membership (based only on their 
kinematics, see LS10) in young moving groups has been complemented with the information provided by 
the equivalent width of the \ion{Li}{i} 6707.8 \AA.\,The lithium equivalent widths are given in 
Table~\ref{tab:upper_branch} and were taken from LS10 and \citet{2010A&A...521A..12M}.  
The latter is an indicator of age for young ($age \lessapprox 650$ Myr) stars
\citep[see details in][]{2001A&A...379..976M,2006ApJ...643.1160L}. For details about the young 
moving groups' ages and the technique used to assign a star to a moving group, see LS10. We note that 
in most cases there is a good agreement between the kinematic classification and the ages obtained 
using lithium. 
Five of these stars (BD+17 232, LQ Hya, V383 Lac, V889 Her, and BD+17 4799) have values of 
$EW\mathrm(\ion{Li}{i})$ higher than those found for the Pleiades members. The remaining 
eight stars have equivalent widths compatible with having an age similar to the Pleiades cluster. 
Finally,
two stars, DK Leo and FP Cnc, were assigned to the Local Association moving group 
but given that they are late type stars, their age could not be confirmed by means of lithium dating (LS10). 
In addition the star 2RE J1846+191 was classified as a YD star by means of its kinematics but 
its age could not be confirmed by lithium dating (LS10).

Therefore, from the 39 stars in the upper branch,  30 are younger 
than the Pleiades or have an age similar to that of the latter cluster, confirming the 
hypothesis of the saturated X-ray regime's corresponding to young stars.

The remaining nine stars in the upper branch in Fig.~\ref{fig:gap} are dwarf flare 
(UV Ceti type) stars (see Table~\ref{tab:upper_branch}). We have classified 
a star as flare when it was included in \citet{1999A&AS..139..555G} and/or \cite{1991MmSAI..62..217P}. 
Some of them have been previously classified as members of a young moving group 
\citep[][LS10]{2001MNRAS.328...45M}
 although they need spectroscopic confirmation.
Nevertheless, whatever their age is, they present enhanced X-ray emission.
They, in fact, present X-ray emission levels in the saturation regime.

\subsubsection{Flare stars}

A large fraction (41\%) of the stars in the upper branch (see Sect.~\ref{sec:peculiarities_M})
are well-known M UV Ceti-type flare dwarfs (see Table~\ref{tab:upper_branch}) or K-type  
flare stars (PW And, BD+17 232, BD+20 1790, DK Leo, and FP Cnc) discovered by \citet{2003A&A...411..489L} and \citet{2005ESASP.560..825M}. 
It is well known that during a flare event, and given that 
Calcium and Balmer lines form at different heights in the stellar atmosphere, the growth of the latter are larger than that 
of the former \citep{2003ApJ...597..535H,2004Ap&SS.292..697C,2006A&A...452..987C}.  
This fact suggests the idea that flare-like events may be responsible for the different behaviour that these stars show
in the flux--flux relationship. So, the question arises: are flare-induced
variations compatible with our observations?

\cite{2004Ap&SS.292..697C,2006A&A...452..987C} carried out a high temporal resolution
spectroscopic monitoring of the stars V1054 Oph and AD Leo, respectively.
Studying the temporal evolution of the Balmer series and \ion{Ca}{ii}~H \& K lines,
they detected frequent ($> 0.71$ flares/hour) short and weak non white-light flares.
Some of these events lasted as little as 14 min. Additional observations by
\cite{2005ESASP.560..491C} suggest that this kind of frequent short flare is
of common occurence on UV Ceti-type flare stars. In the present work, the observed spectra
were obtained using typical exposure times $\sim 25$ min. Thus, one or two of this
kind of flare events probably occurred on the flare stars during their observation.
However, even if this were the case, it would not explain the different behaviour
found in the flux--flux relationships between the stars in the upper branch and
the other ones. In fact, \cite{2006A&A...452..987C} derived (for AD Leo) a value of the order
of $10^{30}$\,erg for the energy released in the $H\alpha$ line by a flare event of the
above type with a duration of 30 min. This implies that flares of this type would contribute with a mean
$H\alpha$ surface flux of the order of $10^5$\,erg~cm$^{-2}$~s$^{-1}$. Thus, one,
or even two, of these events taking place on some of the target
stars during our observations would only introduce a small increase in their measured
line fluxes, which would not explain the whole observed `deviation' of the stars
in the upper branch from the general trend followed by the remainder.
Moreover, the clear trend followed by the stars of the upper branch in the
flux--flux relationship cannot be the result of events of this type occurring randomly in
their atmospheres during the time they were observed.

However, we have detected emission in the \ion{He}{i} D$_{\rm 3}$ line in all the stars classified 
as flare stars located in the upper branch. In active stars, the \ion{He}{i}
D$_{\rm 3}$ line is usually observed in absorption and sometimes in emission, as during
flare events \citep{1997A&AS..125..263M,1999MNRAS.305...45M,2003A&A...411..489L,
2003A&A...397..285G,2006A&A...452..987C}. The \ion{He}{i} D$_{\rm 3}$ line at
$\lambda$5876 \AA\, has a very high excitation level. High temperatures ($> 15000$~K) 
and electron densities $> 10^{\rm 14}$~cm$^{\rm -3}$ are required to reach emission
in this line \citep{1983ApJ...271..832F}. Nanoflare heating has been proposed as the main heating mechanism 
of the stellar outer atmospheres. The energy distribution of flares has been found to be a power law
\citep{1974SoPh...39..155D,1984ApJ...283..421L} of the form\footnote{In this
power law, $dN$ is the number of flares (per unit time) with a total energy
(thermal or radiated) in the interval [$E$,~$E+dE$], and $\alpha$ is greater than 0
\citep[see Sect.~1 in][for the value of $\alpha$ derived by different authors and
its relation with nanoflare heating of quiescent coronae]{2007A&A...471..929C}.} ${dN}/{dE}=kE^{-\alpha}$.
\cite{2006A&A...452..987C} concluded that very weak flares are expected to occur
much more frequently than those observed in that work ($> 0.71$ flares/hour), 
in the sense that the quiescent emission of the UV Ceti-type
stars in the upper branch may be the result of a superposition of multiple small
flares (named nanoflares) following the distribution law given above.

\section{Summary and conclusions}

In this work, we present empirical flux--flux power law relationships between 
the most important chromospheric activity indicators for a large sample of main-sequence  
F, G, K, and M stars. In addition, we obtain relations between 
X-ray and chromospheric surface fluxes for different chromospheric lines. 
These new relationships will be useful for comparing classical and present/future
data of magnetic activity.

For the first time, we have proved the non-universality of some flux--flux relations between chromospheric 
indicators. General power laws hold for the majority of the stars, but some late-K and M dwarf stars deviate 
from the general trend when H$\alpha$ is used as 
a chromospheric activity diagnostic. We have also confirmed that this different 
behaviour persists when X-ray fluxes are used instead of H$\alpha$ ones. Therefore, 
late-type stars follow two different power law flux--flux relationships when X-rays or H$\alpha$ 
are used as magnetic activity indicators. We have quantified, for the first time, 
the departure of these stars from the general trend of other active stars. 
>From their membership in young stellar associations or moving groups
and/or lithium abundance, we have shown that most of the stars 
in the upper branch of the above mentioned relationships are indeed 
young stars. This is also confirmed by their position in the 
$\log F_\mathrm{H\alpha}$--($B$--$V$) diagram.

The remaining stars in the upper branch are flare stars. We have proved 
that a single flare event occurring during the observation cannot account for the deviation observed in the 
flux--flux relationships. However, a plausible explanation for this deviation may be the 
hypothesis of nanoflare heating, which suggests that the quiescent state of the star is 
the result of the superposition of multiple small flares.

Regardless of whether the explanation is the age of the stars or such nanoflare heating, and given that 
all the upper branch stars are in the (magnetic activity) saturation regime, it is our belief that 
their  different behaviour in the flux--flux relationships is a consequence of a shared physical phenomenon. 
These stars probably have a magnetic structure that differs in some way from that of the less active 
stars.

\section*{Acknowledgements}
The authors acknowledge support from the Spanish Ministerio de Educaci\'on y 
Ciencia (currently the Ministerio de Ciencia e Innovaci\'on), under the grant 
FPI20061465-00592 (Programa Nacional Formaci\'on Personal Investigador) and projects 
AYA2008-00695 (Programa Nacional de Astronom\'ia y Astrof\'isica), AstroMadrid
S2009/ESP-1496. J. L\'opez-Santiago acknowledges support by the Spanish Ministerio de Ciencia 
e Innovaci\'on under grant AYA2008-06423-C03-03.
This research has made use of the SIMBAD database and VizieR 
catalogue access tool, operated at CDS, Strasbourg, France. 

\bibliographystyle{mn2e}
\bibliography{activity_rma_final_clean}

\appendix
\section{Tables of results}

The stellar and line parameters are published in electronic format only. 
Table \ref{tab:parameters}, available at the CDS, contains the name of the star
(column \#1), the right ascension and declination (columns \#2 and \#3),  
the spectral type (column \#4), the colour index ($B$--$V$) (column \#5), and 
any important note on each star (column \#6).\\

The chromospheric activity results are listed in two different tables. 
Table \ref{tab:activity_ew} contains the excess emission EW as measured in the subtracted 
spectrum, whereas Table \ref{tab:activity_flux} includes the excess fluxes derived in this work. 
In both tables, column \#1 is the name of the star. In columns \#2, \#3, \#4, \#5, \#6 and \#7, the 
excess emission (or fluxes) for  Ca \textsc{ii} K, Ca \textsc{ii} H, H$\alpha$ and 
Ca \textsc{ii} IRT $\lambda$8498\AA, Ca \textsc{ii} IRT $\lambda$8662\AA, and \ion{He}{i} 
D$_{\rm 3}$ are given.

\tiny
\include{tabla_1_paper}
\include{tabla_2_paper}
\include{tabla_3_paper}
\normalsize

\end{document}

%% file: tabla_1_paper.tex
\begin{table*}
\caption{Stellar parameters for the FEROS05 stars}
\label{tab:parameters}
\centering
\begin{tabular}{ l l l l l} 
\hline
\noalign{\smallskip}
{Name} & \multicolumn{1}{c}{RA} &  \multicolumn{1}{c}{DEC} &  \multicolumn{1}{c}{SpT} & B-V \\
 & \multicolumn{1}{c}{(hhmmss)} & \multicolumn{1}{c}{(deg)} & & \\
\noalign{\smallskip}
\hline
V1005 Ori	&	 04 59 34.83	&	 +01 47 00.7	&	 M1.5V 	&	1.41	 \\
HIP 23309 	&	 05 00 47.13 	&	 -57 15 25.5	&	M0.5V	&	1.40	 \\
HD 35650 	&	 05 24 30.17	&	 -38 58 10.8	&	 K4V 	&	1.25	 \\
V371 Ori	&	 05 33 44.79	&	 +01 56 43.4	&	 M2.5V 	&	1.58	 \\
UY Pic B 	&	 05 36 55.07	&	 -47 57 47.9	&	 K5V 	&	0.79	 \\
AO Men	&	 06 18 28.21	&	 -72 02 41.5	&	 K4IV 	&	1.13	 \\
HIP 31878 	&	 06 39 50.02	&	 -61 28 41.5	&	 M0V 	&	1.26	 \\
V372 Pup	&	 07 28 51.37	&	 -30 14 48.5	&	 M1.5IV 	&	1.44	 \\
YZ CMi	&	 07 44 40.17	&	 +03 33 08.8	&	 M4.5V 	&	1.61	 \\
FR Cnc	&	 08 32 30.53 	&	 +15 49 26.2	&	 K8V 	&	1.16	 \\
GJ 382 	&	 10 12 17.67	&	 -03 44 44.4	&	 M1.5V 	&	1.51	 \\
EE Leo	&	 10 50 52.06	&	 +06 48 29.3	&	 M3.5V 	&	1.68	 \\
V857 Cen	&	 11 31 46.51	&	 -41 02 47.2	&	 M4.5V	&	1.54	 \\
FI Vir	&	 11 47 44.40	&	 +00 48 16.4	&	 M4V 	&	1.77	 \\
EQ Vir	&	 13 34 43.21	&	 -08 20 31.3	&	 K4V 	&	1.20	 \\
CE Boo	&	 14 54 29.24	&	 +16 06 03.8	&	 M2.5V 	&	1.48	 \\
HD 139751 	&	 15 40 28.39	&	 -18 41 46.2	&	 K4V 	&	1.40	 \\
V2306 Oph	&	 16 30 18.06	&	 -12 39 45.3	&	 M3.5V 	&	1.60	 \\
GJ 643 	&	 16 55 25.22	&	 -08 19 21.3	&	 M4V 	&	1.70	 \\
V1054 Oph	&	 16 55 28.75	&	 -08 20 10.8	&	 M4V 	&	1.49	 \\
GJ 674 	&	 17 28 39.95	&	 -46 53 42.7	&	 M2.5V 	&	1.53	 \\
\noalign{\smallskip}
\hline
\end{tabular}
\end{table*}

%% file: tabla_2_paper.tex
\begin{table*}
\caption{Excess emission in different chromospheric activity indicator lines for the active stars in the sample.
\label{tab:activity_ew}}
\centering
\begin{tabular}{ l c c c c c c c}
\hline
\noalign{\smallskip}
& \multicolumn{5}{c}{EW (\AA) in the subtrated spectrum}\\ \cline{2-7}
{Name} & \multicolumn{2}{c}{Ca \textsc {ii}}
 & & \multicolumn{2}{c}{Ca \textsc {ii} IRT} &\\
\cline{2-3} \cline{5-6} & {K} & {H} & ${H\alpha}$ & ${\lambda8498}$ & ${\lambda8662}$ & \ion{He}{i} D3\\
\noalign{\smallskip}
\hline
V1005 Ori	&	 2.100 $\pm$ 0.230 	&	 1.900 $\pm$ 0.200 	&	 1.940 $\pm$ 0.030 	&	 0.460 $\pm$ 0.010 	&	 0.591 $\pm$ 0.012 	&	 0.054 $\pm$ 0.036 \\
HIP 23309 	&	 4.590 $\pm$ 0.050 	&	 3.600 $\pm$ 0.040 	&	 1.490 $\pm$ 0.020 	&	 0.420 $\pm$ 0.010 	&	 0.401 $\pm$ 0.010 	&	 0.023 $\pm$ 0.014 \\
HD 35650 	&	 1.680 $\pm$ 0.020 	&	 1.350 $\pm$ 0.020 	&	 0.440 $\pm$ 0.010 	&	 0.220 $\pm$ 0.010 	&	 0.273 $\pm$ 0.007 	&	 ... \\
V371 Ori	&	 4.830 $\pm$ 0.200 	&	 3.740 $\pm$ 0.160 	&	 4.150 $\pm$ 0.030 	&	 0.390 $\pm$ 0.010 	&	 0.440 $\pm$ 0.016 	&	 0.305 $\pm$ 0.062 \\
UY Pic B 	&	 3.510 $\pm$ 0.030 	&	 2.790 $\pm$ 0.020 	&	 1.760 $\pm$ 0.010 	&	 0.480 $\pm$ 0.010 	&	 0.636 $\pm$ 0.105 	&	 ... \\
AO Men	&	 1.940 $\pm$ 0.060 	&	 1.840 $\pm$ 0.040 	&	 1.460 $\pm$ 0.020 	&	 0.260 $\pm$ 0.050 	&	 ... 	&	 ... \\
HIP 31878 	&	 1.120 $\pm$ 0.040 	&	 0.940 $\pm$ 0.040 	&	 0.360 $\pm$ 0.010 	&	 0.140 $\pm$ 0.010 	&	 0.160 $\pm$ 0.010 	&	 ... \\
V372 Pup	&	 9.070 $\pm$ 0.040 	&	 6.780 $\pm$ 0.020 	&	 3.080 $\pm$ 0.020 	&	 0.550 $\pm$ 0.010 	&	 0.604 $\pm$ 0.007 	&	 0.239 $\pm$ 0.020 \\
YZ CMi	&	 4.550 $\pm$ 0.200 	&	 5.970 $\pm$ 0.170 	&	 7.790 $\pm$ 0.040 	&	 0.810 $\pm$ 0.080 	&	 ... 	&	 0.757 $\pm$ 0.062 \\
FR Cnc	&	5.835 $\pm$ 0.575	&	5.389 $\pm$0.316	&	1.600 $\pm$ 0.022	&	0.601 $\pm$ 0.048	&	 0.576 $\pm$ 0.090 	&	0.895 $\pm$ 0.385 \\
GJ 382 	&	 0.850 $\pm$ 0.040 	&	 0.620 $\pm$ 0.050 	&	 0.330 $\pm$ 0.020 	&	 0.230 $\pm$ 0.040 	&	 ... 	&	 ... \\
EE Leo	&	 0.730 $\pm$ 0.200 	&	 0.500 $\pm$ 0.100 	&	 0.100 $\pm$ 0.030 	&	 0.050 $\pm$ 0.020 	&	 0.099 $\pm$ 0.013 	&	 ... \\
V857 Cen	&	 ... 	&	 8.360 $\pm$ 0.090 	&	 4.600 $\pm$ 0.030 	&	 0.390 $\pm$ 0.050 	&	 0.273 $\pm$ 0.007 	&	 0.382 $\pm$ 0.021 \\
FI Vir	&	 0.480 $\pm$ 0.150 	&	 0.280 $\pm$ 0.180 	&	 0.200 $\pm$ 0.030 	&	 0.090 $\pm$ 0.020 	&	 0.194 $\pm$ 0.016 	&	 ... \\
EQ Vir	&	 2.090 $\pm$ 0.030 	&	 1.700 $\pm$ 0.020 	&	 1.260 $\pm$ 0.010 	&	 0.410 $\pm$ 0.010 	&	 0.052 $\pm$ 0.011 	&	 0.037 $\pm$ 0.014 \\
CE Boo	&	 4.320 $\pm$ 0.010 	&	 3.280 $\pm$ 0.040 	&	 1.380 $\pm$ 0.010 	&	 0.210 $\pm$ 0.010 	&	 0.304 $\pm$ 0.009 	&	 0.094 $\pm$ 0.018 \\
HD 139751 	&	 2.590 $\pm$ 0.020 	&	 2.030 $\pm$ 0.030 	&	 1.300 $\pm$ 0.020 	&	 0.480 $\pm$ 0.010 	&	 0.519 $\pm$ 0.012 	&	 ... \\
V2306 Oph	&	 0.600 $\pm$ 0.160 	&	 0.360 $\pm$ 0.190 	&	 0.130 $\pm$ 0.020 	&	 0.060 $\pm$ 0.030 	&	 0.070 $\pm$ 0.015 	&	 ... \\
GJ 643 	&	 0.730 $\pm$ 0.240 	&	 0.290 $\pm$ 0.130 	&	 0.100 $\pm$ 0.010 	&	 0.050 $\pm$ 0.030 	&	 0.075 $\pm$ 0.008 	&	 ... \\
V1054 Oph	&	 5.080 $\pm$ 0.070 	&	 3.070 $\pm$ 0.040 	&	 2.780 $\pm$ 0.030 	&	 0.170 $\pm$ 0.020 	&	 0.180 $\pm$ 0.007 	&	 0.111 $\pm$ 0.034 \\
GJ 674 	&	 0.340 $\pm$ 0.040 	&	 0.140 $\pm$ 0.040 	&	 0.260 $\pm$ 0.020 	&	 ... 	&	 0.099 $\pm$ 0.009 	&	 ... \\
\noalign{\smallskip}
\hline
\end{tabular}
\end{table*}

%% file: tabla_3_paper.tex
\begin{table*}
\caption{Logarithmic excess surface flux in different chromospheric activity indicator lines for the active stars in the sample.
\label{tab:activity_flux}}
\centering
\begin{tabular}{ l c c c c c c c}
\hline
\noalign{\smallskip}
& \multicolumn{5}{c}{$\log{F_{\rm s}}$ in the subtrated spectrum}\\ \cline{2-7}
{Name} & \multicolumn{2}{c}{Ca \textsc {ii}}
 & & \multicolumn{2}{c}{Ca \textsc {ii} IRT} &\\
\cline{2-3} \cline{5-6} & {K} & {H} & ${H\alpha}$ & ${\lambda8498}$ & ${\lambda8662}$\\
\noalign{\smallskip}
\hline
V1005 Ori	&	 5.730 $\pm$ 0.048 	&	 5.687 $\pm$ 0.048 	&	 6.296 $\pm$ 0.020 	&	 5.780 $\pm$ 0.019 	&	 5.889 $\pm$ 0.018 \\
HIP 23309 	&	 6.090 $\pm$ 0.005 	&	 5.984 $\pm$ 0.005 	&	 6.192 $\pm$ 0.017 	&	 5.747 $\pm$ 0.021 	&	 5.727 $\pm$ 0.021 \\
HD 35650 	&	 5.953 $\pm$ 0.005 	&	 5.858 $\pm$ 0.005 	&	 5.825 $\pm$ 0.030 	&	 5.569 $\pm$ 0.039 	&	 5.663 $\pm$ 0.021 \\
V371 Ori	&	 5.753 $\pm$ 0.018 	&	 5.642 $\pm$ 0.018 	&	 6.442 $\pm$ 0.009 	&	 5.592 $\pm$ 0.022 	&	 5.645 $\pm$ 0.031 \\
UY Pic B 	&	 7.190 $\pm$ 0.004 	&	 7.091 $\pm$ 0.004 	&	 6.926 $\pm$ 0.007 	&	 6.223 $\pm$ 0.018 	&	 6.345 $\pm$ 0.144 \\
AO Men	&	 6.254 $\pm$ 0.013 	&	 6.231 $\pm$ 0.013 	&	 6.476 $\pm$ 0.018 	&	 5.724 $\pm$ 0.167 	&	 ... \\
HIP 31878 	&	 5.757 $\pm$ 0.016 	&	 5.680 $\pm$ 0.016 	&	 5.727 $\pm$ 0.036 	&	 5.366 $\pm$ 0.062 	&	 5.425 $\pm$ 0.055 \\
V372 Pup	&	 6.306 $\pm$ 0.002 	&	 6.179 $\pm$ 0.002 	&	 6.464 $\pm$ 0.008 	&	 5.837 $\pm$ 0.016 	&	 5.878 $\pm$ 0.011 \\
YZ CMi	&	 5.667 $\pm$ 0.019 	&	 5.785 $\pm$ 0.019 	&	 6.683 $\pm$ 0.007 	&	 5.889 $\pm$ 0.086 	&	 ... \\
FR Cnc	&	 5.141 $\pm$ 0.015 	&	 5.173 $\pm$ 0.012 	&	 6.072 $\pm$ 0.007 	&	 6.513 $\pm$ 0.005 	&	 6.519 $\pm$ 0.006 \\ 
GJ 382 	&	 5.138 $\pm$ 0.020 	&	 5.001 $\pm$ 0.020 	&	 5.418 $\pm$ 0.079 	&	 5.410 $\pm$ 0.151 	&	 ... \\
EE Leo	&	 4.733 $\pm$ 0.119 	&	 4.568 $\pm$ 0.119 	&	 4.715 $\pm$ 0.391 	&	 4.631 $\pm$ 0.347 	&	 4.929 $\pm$ 0.116 \\
V857 Cen	&	 5.432 $\pm$ 0.006 	&	 5.293 $\pm$ 0.006 	&	 6.107 $\pm$ 0.008 	&	 5.352 $\pm$ 0.111 	&	 5.196 $\pm$ 0.022 \\
FI Vir	&	 4.371 $\pm$ 0.136 	&	 4.137 $\pm$ 0.136 	&	 4.919 $\pm$ 0.195 	&	 4.825 $\pm$ 0.193 	&	 5.158 $\pm$ 0.070 \\
EQ Vir 	&	 6.147 $\pm$ 0.006 	&	 6.057 $\pm$ 0.006 	&	 6.336 $\pm$ 0.010 	&	 5.874 $\pm$ 0.021 	&	 4.974 $\pm$ 0.190 \\
CE Boo	&	 5.904 $\pm$ 0.001 	&	 5.784 $\pm$ 0.001 	&	 6.072 $\pm$ 0.009 	&	 5.391 $\pm$ 0.041 	&	 5.552 $\pm$ 0.024 \\
HD 139751 	&	 5.841 $\pm$ 0.003 	&	 5.735 $\pm$ 0.003 	&	 6.133 $\pm$ 0.020 	&	 5.805 $\pm$ 0.018 	&	 5.840 $\pm$ 0.019 \\
V2306 Oph	&	 4.807 $\pm$ 0.116 	&	 4.585 $\pm$ 0.116 	&	 4.916 $\pm$ 0.200 	&	 4.765 $\pm$ 0.434 	&	 4.830 $\pm$ 0.186 \\
GJ 643 	&	 4.693 $\pm$ 0.143 	&	 4.292 $\pm$ 0.143 	&	 4.694 $\pm$ 0.130 	&	 4.617 $\pm$ 0.521 	&	 4.793 $\pm$ 0.095 \\
V1054 Oph	&	 5.954 $\pm$ 0.006 	&	 5.736 $\pm$ 0.006 	&	 6.365 $\pm$ 0.014 	&	 5.293 $\pm$ 0.102 	&	 5.318 $\pm$ 0.034 \\
GJ 674 	&	 4.700 $\pm$ 0.051 	&	 4.315 $\pm$ 0.051 	&	 5.293 $\pm$ 0.100 	&	 ... 	&	 5.029 $\pm$ 0.082 \\
\noalign{\smallskip}
\hline
\end{tabular}
\end{table*}

%% file: martinez_arnaiz_fluxes_revised.bbl
\begin{thebibliography}{}

\bibitem[\protect\citeauthoryear{{Baliunas}, {Sokoloff} \& {Soon}}{{Baliunas}
  et~al.}{1996}]{1996ApJ...457L..99B}
{Baliunas} S.,  {Sokoloff} D.,    {Soon} W.,  1996, ApJ Letters, 457, L99

\bibitem[\protect\citeauthoryear{{Barden}}{{Barden}}{1985}]{1985ApJ...295..162%
B}
{Barden} S.~C.,  1985, ApJ, 295, 162

\bibitem[\protect\citeauthoryear{{Barnes}}{{Barnes}}{2003}]{2003ApJ...586..464%
B}
{Barnes} S.~A.,  2003, ApJ, 586, 464

\bibitem[\protect\citeauthoryear{{Barrado y Navascu{\'e}s} \&
  {Mart{\'{\i}}n}}{{Barrado y Navascu{\'e}s} \&
  {Mart{\'{\i}}n}}{2003}]{2003AJ....126.2997B}
{Barrado y Navascu{\'e}s} D.,  {Mart{\'{\i}}n} E.~L.,  2003, AJ, 126, 2997

\bibitem[\protect\citeauthoryear{{Barrado y Navascu{\'e}s}, {Stauffer}, {Song}
  \& {Caillault}}{{Barrado y Navascu{\'e}s} et~al.}{1999}]{1999ApJ...520L.123B}
{Barrado y Navascu{\'e}s} D.,  {Stauffer} J.~R.,  {Song} I.,    {Caillault} J.,
   1999, ApJ Letters, 520, L123

\bibitem[\protect\citeauthoryear{{B{\"o}hm-Vitense}}{{B{\"o}hm-Vitense}}{2007}%
]{2007ApJ...657..486B}
{B{\"o}hm-Vitense} E.,  2007, ApJ, 657, 486

\bibitem[\protect\citeauthoryear{{Bopp} \& {Fekel} Jr.}{{Bopp} \&
  {Fekel}}{1977}]{1977AJ.....82..490B}
{Bopp} B.~W.,  {Fekel} Jr. F.,  1977, AJ, 82, 490

\bibitem[\protect\citeauthoryear{{Bus{\`a}}, {Aznar Cuadrado}, {Terranegra},
  {Andretta} \& {Gomez}}{{Bus{\`a}} et~al.}{2007}]{2007A&A...466.1089B}
{Bus{\`a}} I.,  {Aznar Cuadrado} R.,  {Terranegra} L.,  {Andretta} V.,
  {Gomez} M.~T.,  2007, A\&A, 466, 1089

\bibitem[\protect\citeauthoryear{{Cincunegui}, {D{\'{\i}}az} \&
  {Mauas}}{{Cincunegui} et~al.}{2007}]{2007A&A...469..309C}
{Cincunegui} C.,  {D{\'{\i}}az} R.~F.,    {Mauas} P.~J.~D.,  2007, A\&A, 469,
  309

\bibitem[\protect\citeauthoryear{{Crespo-Chac{\'o}n}, {Micela}, {Reale},
  {Caramazza}, {L{\'o}pez-Santiago} \& {Pillitteri}}{{Crespo-Chac{\'o}n}
  et~al.}{2007}]{2007A&A...471..929C}
{Crespo-Chac{\'o}n} I.,  {Micela} G.,  {Reale} F.,  {Caramazza} M.,
  {L{\'o}pez-Santiago} J.,    {Pillitteri} I.,  2007, A\&A, 471, 929

\bibitem[\protect\citeauthoryear{{Crespo-Chac{\'o}n}, {Montes},
  {Fern{\'a}ndez-Figueroa} \& {L{\'o}pez-Santiago}}{{Crespo-Chac{\'o}n}
  et~al.}{2005}]{2005ESASP.560..491C}
{Crespo-Chac{\'o}n} I.,  {Montes} D.,  {Fern{\'a}ndez-Figueroa} M.~J.,
  {L{\'o}pez-Santiago} J.,  2005, in {F.~Favata, G.~A.~J.~Hussain, \&
  B.~Battrick} ed., 13th Cambridge Workshop on Cool Stars, Stellar Systems and
  the Sun Vol.~560 of ESA Special Publication, {High temporal resolution
  spectroscopic observations of UV Cet type flare stars}.
p.~491

\bibitem[\protect\citeauthoryear{{Crespo-Chac{\'o}n}, {Montes},
  {Fern{\'a}ndez-Figueroa}, {L{\'o}pez-Santiago}, {Garc{\'{\i}}a-Alvarez} \&
  {Foing}}{{Crespo-Chac{\'o}n} et~al.}{2004}]{2004Ap&SS.292..697C}
{Crespo-Chac{\'o}n} I.,  {Montes} D.,  {Fern{\'a}ndez-Figueroa} M.~J.,
  {L{\'o}pez-Santiago} J.,  {Garc{\'{\i}}a-Alvarez} D.,    {Foing} B.~H.,
  2004, Astrophysics and Space Science, 292, 697

\bibitem[\protect\citeauthoryear{{Crespo-Chac{\'o}n}, {Montes},
  {Garc{\'{\i}}a-Alvarez}, {Fern{\'a}ndez-Figueroa}, {L{\'o}pez-Santiago} \&
  {Foing}}{{Crespo-Chac{\'o}n} et~al.}{2006}]{2006A&A...452..987C}
{Crespo-Chac{\'o}n} I.,  {Montes} D.,  {Garc{\'{\i}}a-Alvarez} D.,
  {Fern{\'a}ndez-Figueroa} M.~J.,  {L{\'o}pez-Santiago} J.,    {Foing} B.~H.,
  2006, A\&A, 452, 987

\bibitem[\protect\citeauthoryear{{Cuntz}, {Saar} \& {Musielak}}{{Cuntz}
  et~al.}{2000}]{2000ApJ...533L.151C}
{Cuntz} M.,  {Saar} S.~H.,    {Musielak} Z.~E.,  2000, ApJ Letters, 533, L151

\bibitem[\protect\citeauthoryear{{Cutispoto}, {Pastori}, {Tagliaferri},
  {Messina} \& {Pallavicini}}{{Cutispoto} et~al.}{1999}]{1999A&AS..138...87C}
{Cutispoto} G.,  {Pastori} L.,  {Tagliaferri} G.,  {Messina} S.,
  {Pallavicini} R.,  1999, A\&A Suppl., 138, 87

\bibitem[\protect\citeauthoryear{{Datlowe}, {Elcan} \& {Hudson}}{{Datlowe}
  et~al.}{1974}]{1974SoPh...39..155D}
{Datlowe} D.~W.,  {Elcan} M.~J.,    {Hudson} H.~S.,  1974, Solar Physics, 39,
  155

\bibitem[\protect\citeauthoryear{{Durney}, {Mihalas} \& {Robinson}}{{Durney}
  et~al.}{1981}]{1981PASP...93..537D}
{Durney} B.~R.,  {Mihalas} D.,    {Robinson} R.~D.,  1981, Publications of the
  ASP, 93, 537

\bibitem[\protect\citeauthoryear{{Feldman}, {Liggett} \& {Zirin}}{{Feldman}
  et~al.}{1983}]{1983ApJ...271..832F}
{Feldman} U.,  {Liggett} M.,    {Zirin} H.,  1983, ApJ, 271, 832

\bibitem[\protect\citeauthoryear{{Garc{\'{\i}}a-Alvarez}, {Foing}, {Montes},
  {Oliveira}, {Doyle}, {Messina} \& {et al.}}{{Garc{\'{\i}}a-Alvarez}
  et~al.}{2003}]{2003A&A...397..285G}
{Garc{\'{\i}}a-Alvarez} D.,  {Foing} B.~H.,  {Montes} D.,  {Oliveira} J.,
  {Doyle} J.~G.,  {Messina} S.,    {et al.} 2003, A\&A, 397, 285

\bibitem[\protect\citeauthoryear{{Gershberg}, {Katsova}, {Lovkaya}, {Terebizh}
  \& {Shakhovskaya}}{{Gershberg} et~al.}{1999}]{1999A&AS..139..555G}
{Gershberg} R.~E.,  {Katsova} M.~M.,  {Lovkaya} M.~N.,  {Terebizh} A.~V.,
  {Shakhovskaya} N.~I.,  1999, A\&A Suppl., 139, 555

\bibitem[\protect\citeauthoryear{{Hall}}{{Hall}}{1996}]{1996PASP..108..313H}
{Hall} J.~C.,  1996, Publications of the ASP, 108, 313

\bibitem[\protect\citeauthoryear{{Hartmann}, {Soderblom}, {Noyes}, {Burnham} \&
  {Vaughan}}{{Hartmann} et~al.}{1984}]{1984ApJ...276..254H}
{Hartmann} L.,  {Soderblom} D.~R.,  {Noyes} R.~W.,  {Burnham} N.,    {Vaughan}
  A.~H.,  1984, ApJ, 276, 254

\bibitem[\protect\citeauthoryear{{Hartmann} \& {Noyes}}{{Hartmann} \&
  {Noyes}}{1987}]{1987ARA&A..25..271H}
{Hartmann} L.~W.,  {Noyes} R.~W.,  1987, A\&A Annual Review, 25, 271

\bibitem[\protect\citeauthoryear{{Hawley}, {Allred}, {Johns-Krull}, {Fisher},
  {Abbett}, {Alekseev}, {Avgoloupis}, {Deustua}, {Gunn}, {Seiradakis}, {Sirk}
  \& {Valenti}}{{Hawley} et~al.}{2003}]{2003ApJ...597..535H}
{Hawley} S.~L.,  {Allred} J.~C.,  {Johns-Krull} C.~M.,  {Fisher} G.~H.,
  {Abbett} W.~P.,  {Alekseev} I.,  {Avgoloupis} S.~I.,  {Deustua} S.~E.,
  {Gunn} A.,  {Seiradakis} J.~H.,  {Sirk} M.~M.,    {Valenti} J.~A.,  2003,
  ApJ, 597, 535

\bibitem[\protect\citeauthoryear{{Hempelmann}, {Schmitt}, {Schultz}, {Ruediger}
  \& {Stepien}}{{Hempelmann} et~al.}{1995}]{1995A&A...294..515H}
{Hempelmann} A.,  {Schmitt} J.~H.~M.~M.,  {Schultz} M.,  {Ruediger} G.,
  {Stepien} K.,  1995, A\&A, 294, 515

\bibitem[\protect\citeauthoryear{{Henry}, {Baliunas}, {Donahue}, {Soon} \&
  {Saar}}{{Henry} et~al.}{1997}]{1997ApJ...474..503H}
{Henry} G.~W.,  {Baliunas} S.~L.,  {Donahue} R.~A.,  {Soon} W.~H.,    {Saar}
  S.~H.,  1997, ApJ, 474, 503

\bibitem[\protect\citeauthoryear{{Hern{\'a}n-Obispo}, {G{\'a}lvez-Ortiz},
  {Anglada-Escud{\'e}}, {Kane}, {Barnes}, {de Castro} \&
  {Cornide}}{{Hern{\'a}n-Obispo} et~al.}{2010}]{2010A&A...512A..45H}
{Hern{\'a}n-Obispo} M.,  {G{\'a}lvez-Ortiz} M.~C.,  {Anglada-Escud{\'e}} G.,
  {Kane} S.~R.,  {Barnes} J.~R.,  {de Castro} E.,    {Cornide} M.,  2010, A\&A,
  512, A45

\bibitem[\protect\citeauthoryear{{Hooten} \& {Hall}}{{Hooten} \&
  {Hall}}{1990}]{1990ApJS...74..225H}
{Hooten} J.~T.,  {Hall} D.~S.,  1990, ApJ Suppl., 74, 225

\bibitem[\protect\citeauthoryear{{Huenemoerder} \& {Barden}}{{Huenemoerder} \&
  {Barden}}{1984}]{1984BAAS...16..510H}
{Huenemoerder} D.~P.,  {Barden} S.~C.,  1984, in Bulletin of the American
  Astronomical Society Vol.~16, {Spectroscopy of RS CVn and W UMa Systems:
  Extraction of Component Spectra}.
p.~510

\bibitem[\protect\citeauthoryear{{Isobe}, {Feigelson}, {Akritas} \&
  {Babu}}{{Isobe} et~al.}{1990}]{1990ApJ...364..104I}
{Isobe} T.,  {Feigelson} E.~D.,  {Akritas} M.~G.,    {Babu} G.~J.,  1990, ApJ,
  364, 104

\bibitem[\protect\citeauthoryear{{James}, {Jardine}, {Jeffries}, {Randich},
  {Collier Cameron} \& {Ferreira}}{{James} et~al.}{2000}]{2000MNRAS.318.1217J}
{James} D.~J.,  {Jardine} M.~M.,  {Jeffries} R.~D.,  {Randich} S.,  {Collier
  Cameron} A.,    {Ferreira} M.,  2000, Monthly Notices of the RAS, 318, 1217

\bibitem[\protect\citeauthoryear{{J{\"a}rvinen}, {Korhonen}, {Berdyugina},
  {Ilyin}, {Strassmeier}, {Weber}, {Savanov} \& {Tuominen}}{{J{\"a}rvinen}
  et~al.}{2008}]{2008A&A...488.1047J}
{J{\"a}rvinen} S.~P.,  {Korhonen} H.,  {Berdyugina} S.~V.,  {Ilyin} I.,
  {Strassmeier} K.~G.,  {Weber} M.,  {Savanov} I.,    {Tuominen} I.,  2008,
  A\&A, 488, 1047

\bibitem[\protect\citeauthoryear{{Jenkins}, {Ramsey}, {Jones}, {Pavlenko},
  {Gallardo}, {Barnes} \& {Pinfield}}{{Jenkins}
  et~al.}{2009}]{2009ApJ...704..975J}
{Jenkins} J.~S.,  {Ramsey} L.~W.,  {Jones} H.~R.~A.,  {Pavlenko} Y.,
  {Gallardo} J.,  {Barnes} J.~R.,    {Pinfield} D.~J.,  2009, ApJ, 704, 975

\bibitem[\protect\citeauthoryear{{Koen} \& {Eyer}}{{Koen} \&
  {Eyer}}{2002}]{2002MNRAS.331...45K}
{Koen} C.,  {Eyer} L.,  2002, Monthly Notices of the RAS, 331, 45

\bibitem[\protect\citeauthoryear{{Kov{\'a}ri}, {Strassmeier}, {Granzer},
  {Weber}, {Ol{\'a}h} \& {Rice}}{{Kov{\'a}ri}
  et~al.}{2004}]{2004A&A...417.1047K}
{Kov{\'a}ri} Z.,  {Strassmeier} K.~G.,  {Granzer} T.,  {Weber} M.,  {Ol{\'a}h}
  K.,    {Rice} J.~B.,  2004, A\&A, 417, 1047

\bibitem[\protect\citeauthoryear{{Lin}, {Schwartz}, {Kane}, {Pelling} \&
  {Hurley}}{{Lin} et~al.}{1984}]{1984ApJ...283..421L}
{Lin} R.~P.,  {Schwartz} R.~A.,  {Kane} S.~R.,  {Pelling} R.~M.,    {Hurley}
  K.~C.,  1984, ApJ, 283, 421

\bibitem[\protect\citeauthoryear{{Lister}, {Collier Cameron} \&
  {Bartus}}{{Lister} et~al.}{1999}]{1999MNRAS.307..685L}
{Lister} T.~A.,  {Collier Cameron} A.,    {Bartus} J.,  1999, Monthly Notices
  of the RAS, 307, 685

\bibitem[\protect\citeauthoryear{{L\'opez-Santiago}}{{L\'opez-Santiago}}{2005}%
]{2005PhDT........14B}
{L\'opez-Santiago} J.,  2005, PhD thesis, Universidad Complutense de Madrid

\bibitem[\protect\citeauthoryear{{L{\'o}pez-Santiago}, {Micela} \&
  {Montes}}{{L{\'o}pez-Santiago} et~al.}{2009}]{2009A&A...499..129L}
{L{\'o}pez-Santiago} J.,  {Micela} G.,    {Montes} D.,  2009, A\&A, 499, 129

\bibitem[\protect\citeauthoryear{{L{\'o}pez-Santiago}, {Montes},
  {Crespo-Chac{\'o}n} \& {Fern{\'a}ndez-Figueroa}}{{L{\'o}pez-Santiago}
  et~al.}{2006}]{2006ApJ...643.1160L}
{L{\'o}pez-Santiago} J.,  {Montes} D.,  {Crespo-Chac{\'o}n} I.,
  {Fern{\'a}ndez-Figueroa} M.~J.,  2006, ApJ, 643, 1160

\bibitem[\protect\citeauthoryear{{L{\'o}pez-Santiago}, {Montes},
  {Fern{\'a}ndez-Figueroa}, {G{\'a}lvez} \&
  {Crespo-Chac{\'o}n}}{{L{\'o}pez-Santiago} et~al.}{2005}]{2005ESASP.560..775L}
{L{\'o}pez-Santiago} J.,  {Montes} D.,  {Fern{\'a}ndez-Figueroa} M.~J.,
  {G{\'a}lvez} M.~C.,    {Crespo-Chac{\'o}n} I.,  2005, in {F.~Favata,
  G.~A.~J.~Hussain, \& B.~Battrick} ed., 13th Cambridge Workshop on Cool Stars,
  Stellar Systems and the Sun Vol.~560 of ESA Special Publication, {A study on
  the flux--flux and activity--rotation relationships for late-type stars
  members of young stellar kinematic groups}.
p.~775

\bibitem[\protect\citeauthoryear{{L{\'o}pez-Santiago}, {Montes},
  {Fern{\'a}ndez-Figueroa} \& {Ramsey}}{{L{\'o}pez-Santiago}
  et~al.}{2003}]{2003A&A...411..489L}
{L{\'o}pez-Santiago} J.,  {Montes} D.,  {Fern{\'a}ndez-Figueroa} M.~J.,
  {Ramsey} L.~W.,  2003, A\&A, 411, 489

\bibitem[\protect\citeauthoryear{{L{\'o}pez-Santiago}, {Montes},
  {G{\'a}lvez-Ortiz}, {Crespo-Chac{\'o}n}, {Mart{\'{\i}}nez-Arn{\'a}iz},
  {Fern{\'a}ndez-Figueroa}, {de Castro} \& {Cornide}}{{L{\'o}pez-Santiago}
  et~al.}{2010}]{2010A&A...514A..97L}
{L{\'o}pez-Santiago} J.,  {Montes} D.,  {G{\'a}lvez-Ortiz} M.~C.,
  {Crespo-Chac{\'o}n} I.,  {Mart{\'{\i}}nez-Arn{\'a}iz} R.~M.,
  {Fern{\'a}ndez-Figueroa} M.~J.,  {de Castro} E.,    {Cornide} M.,  2010,
  A\&A, 514, A97

\bibitem[\protect\citeauthoryear{{Maldonado}, {Mart{\'{\i}}nez-Arn{\'a}iz},
  {Eiroa}, {Montes} \& {Montesinos}}{{Maldonado}
  et~al.}{2010}]{2010A&A...521A..12M}
{Maldonado} J.,  {Mart{\'{\i}}nez-Arn{\'a}iz} R.~M.,  {Eiroa} C.,  {Montes} D.,
     {Montesinos} B.,  2010, A\&A, 521, A12

\bibitem[\protect\citeauthoryear{{Mamajek} \& {Hillenbrand}}{{Mamajek} \&
  {Hillenbrand}}{2008}]{2008ApJ...687.1264M}
{Mamajek} E.~E.,  {Hillenbrand} L.~A.,  2008, ApJ, 687, 1264

\bibitem[\protect\citeauthoryear{{Mart{\'{\i}}nez-Arn{\'a}iz}, {Maldonado},
  {Montes}, {Eiroa} \& {Montesinos}}{{Mart{\'{\i}}nez-Arn{\'a}iz}
  et~al.}{2010}]{2010A&A...520A..79M}
{Mart{\'{\i}}nez-Arn{\'a}iz} R.,  {Maldonado} J.,  {Montes} D.,  {Eiroa} C.,
  {Montesinos} B.,  2010, A\&A, 520, A79

\bibitem[\protect\citeauthoryear{{Messina}, {Desidera}, {Turatto}, {Lanzafame}
  \& {Guinan}}{{Messina} et~al.}{2010}]{2010A&A...520A..15M}
{Messina} S.,  {Desidera} S.,  {Turatto} M.,  {Lanzafame} A.~C.,    {Guinan}
  E.~F.,  2010, A\&A, 520, A15

\bibitem[\protect\citeauthoryear{{Messina}, {Pizzolato}, {Guinan} \&
  {Rodon{\`o}}}{{Messina} et~al.}{2003}]{2003A&A...410..671M}
{Messina} S.,  {Pizzolato} N.,  {Guinan} E.~F.,    {Rodon{\`o}} M.,  2003,
  A\&A, 410, 671

\bibitem[\protect\citeauthoryear{{Messina}, {Rodon{\`o}} \& {Guinan}}{{Messina}
  et~al.}{2001}]{2001A&A...366..215M}
{Messina} S.,  {Rodon{\`o}} M.,    {Guinan} E.~F.,  2001, A\&A, 366, 215

\bibitem[\protect\citeauthoryear{{Mohanty} \& {Basri}}{{Mohanty} \&
  {Basri}}{2003}]{2003ApJ...583..451M}
{Mohanty} S.,  {Basri} G.,  2003, ApJ, 583, 451

\bibitem[\protect\citeauthoryear{{Montes}, {de Castro}, {Fern\'andez-Figueroa}
  \& {Cornide}}{{Montes} et~al.}{1995}]{1995A&AS..114..287M}
{Montes} D.,  {de Castro} E.,  {Fern\'andez-Figueroa} M.~J.,    {Cornide} M.,
  1995, A\&A Suppl., 114, 287

\bibitem[\protect\citeauthoryear{{Montes}, {Fern\'andez-Figueroa}, {Cornide} \&
  {de Castro}}{{Montes} et~al.}{1996a}]{1996ASPC..109..657M}
{Montes} D.,  {Fern\'andez-Figueroa} M.~J.,  {Cornide} M.,    {de Castro} E.,
  1996a, in {R.~Pallavicini \& A.~K.~Dupree} ed., Cool Stars, Stellar Systems,
  and the Sun Vol.~109 of ASP Conf. Ser., {Flux--flux relations between excess
  H alpha, CA II H and K and H epsilon emissions and other activity indicators
  in chromospherically active binaries}.
p.~657

\bibitem[\protect\citeauthoryear{{Montes}, {Fern\'andez-Figueroa}, {Cornide} \&
  {de Castro}}{{Montes} et~al.}{1996b}]{1996A&A...312..221M}
{Montes} D.,  {Fern\'andez-Figueroa} M.~J.,  {Cornide} M.,    {de Castro} E.,
  1996b, A\&A, 312, 221

\bibitem[\protect\citeauthoryear{{Montes}, {Fern\'andez-Figueroa}, {de Castro}
  \& {Cornide}}{{Montes} et~al.}{1995}]{1995A&A...294..165M}
{Montes} D.,  {Fern\'andez-Figueroa} M.~J.,  {de Castro} E.,    {Cornide} M.,
  1995, A\&A, 294, 165

\bibitem[\protect\citeauthoryear{{Montes}, {Fern{\'a}ndez-Figueroa}, {de
  Castro}, {Cornide}, {Latorre} \& {Sanz-Forcada}}{{Montes}
  et~al.}{2000}]{2000A&AS..146..103M}
{Montes} D.,  {Fern{\'a}ndez-Figueroa} M.~J.,  {de Castro} E.,  {Cornide} M.,
  {Latorre} A.,    {Sanz-Forcada} J.,  2000, A\&A Suppl., 146, 103

\bibitem[\protect\citeauthoryear{{Montes}, {Fern\'andez-Figueroa}, {de Castro}
  \& {Sanz-Forcada}}{{Montes} et~al.}{1997}]{1997A&AS..125..263M}
{Montes} D.,  {Fern\'andez-Figueroa} M.~J.,  {de Castro} E.,    {Sanz-Forcada}
  J.,  1997, A\&A Suppl., 125, 263

\bibitem[\protect\citeauthoryear{{Montes}, {L{\'o}pez-Santiago},
  {Crespo-Chac{\'o}n} \& {Fern{\'a}ndez-Figueroa}}{{Montes}
  et~al.}{2005}]{2005ESASP.560..825M}
{Montes} D.,  {L{\'o}pez-Santiago} J.,  {Crespo-Chac{\'o}n} I.,
  {Fern{\'a}ndez-Figueroa} M.~J.,  2005, in {F.~Favata, G.~A.~J.~Hussain, \&
  B.~Battrick} ed., 13th Cambridge Workshop on Cool Stars, Stellar Systems and
  the Sun Vol.~560 of ESA Special Publication, {Flare stars among K dwarfs
  members of young stellar kinematic groups}.
p.~825

\bibitem[\protect\citeauthoryear{{Montes}, {L\'opez-Santiago},
  {Crespo-Chac{\'o}n}, {Mart\'inez-Arn\'aiz} \& {Maldonado}}{{Montes}
  et~al.}{2007}]{2007seadmg}
{Montes} D.,  {L\'opez-Santiago} J.,  {Crespo-Chac{\'o}n} I.,
  {Mart\'inez-Arn\'aiz} R.,    {Maldonado} J.,  2007, in Highlights of Spanish
  Astrophysics IV Proceedings of the VII Scientific Meeting of the Spanish
  Astronomical Society (SEA), {High resolution spectroscopic analysis of cool
  stars possible members of the AB Doradus moving group}.
Springer

\bibitem[\protect\citeauthoryear{{Montes}, {L\'opez-Santiago},
  {Crespo-Chac{\'o}n}, {Mart\'inez-Arn\'aiz} \& {Maldonado}}{{Montes}
  et~al.}{2008}]{2008cooldmg}
{Montes} D.,  {L\'opez-Santiago} J.,  {Crespo-Chac{\'o}n} I.,
  {Mart\'inez-Arn\'aiz} R.,    {Maldonado} J.,  2008, in Proceedings of The
  14th Cool Stars, Stellar Systems and the Sun workshop ASP conf. Ser., V. 384,
  {High resolution spectroscopic analysis of cool stars possible members of
  nearby young moving groups.}

\bibitem[\protect\citeauthoryear{{Montes}, {L{\'o}pez-Santiago},
  {Fern{\'a}ndez-Figueroa} \& {G{\'a}lvez}}{{Montes}
  et~al.}{2001}]{2001A&A...379..976M}
{Montes} D.,  {L{\'o}pez-Santiago} J.,  {Fern{\'a}ndez-Figueroa} M.~J.,
  {G{\'a}lvez} M.~C.,  2001, A\&A, 379, 976

\bibitem[\protect\citeauthoryear{{Montes}, {L{\'o}pez-Santiago}, {G{\'a}lvez},
  {Fern{\'a}ndez-Figueroa}, {de Castro} \& {Cornide}}{{Montes}
  et~al.}{2001}]{2001MNRAS.328...45M}
{Montes} D.,  {L{\'o}pez-Santiago} J.,  {G{\'a}lvez} M.~C.,
  {Fern{\'a}ndez-Figueroa} M.~J.,  {de Castro} E.,    {Cornide} M.,  2001,
  Monthly Notices of the RAS, 328, 45

\bibitem[\protect\citeauthoryear{{Montes}, {Saar}, {Collier Cameron} \&
  {Unruh}}{{Montes} et~al.}{1999}]{1999MNRAS.305...45M}
{Montes} D.,  {Saar} S.~H.,  {Collier Cameron} A.,    {Unruh} Y.~C.,  1999,
  Monthly Notices of the RAS, 305, 45

\bibitem[\protect\citeauthoryear{{Norton}, {Wheatley}, {West}, {Haswell},
  {Street}, {Collier Cameron} \& {et al.}}{{Norton}
  et~al.}{2007}]{2007A&A...467..785N}
{Norton} A.~J.,  {Wheatley} P.~J.,  {West} R.~G.,  {Haswell} C.~A.,  {Street}
  R.~A.,  {Collier Cameron} A.,    {et al.} 2007, A\&A, 467, 785

\bibitem[\protect\citeauthoryear{{Noyes}, {Hartmann}, {Baliunas}, {Duncan} \&
  {Vaughan}}{{Noyes} et~al.}{1984}]{1984ApJ...279..763N}
{Noyes} R.~W.,  {Hartmann} L.~W.,  {Baliunas} S.~L.,  {Duncan} D.~K.,
  {Vaughan} A.~H.,  1984, ApJ, 279, 763

\bibitem[\protect\citeauthoryear{{Oranje}}{{Oranje}}{1986}]{1986A&A...154..185%
O}
{Oranje} B.~J.,  1986, A\&A, 154, 185

\bibitem[\protect\citeauthoryear{{Pace}, {Melendez}, {Pasquini}, {Carraro},
  {Danziger}, {Fran{\c c}ois}, {Matteucci} \& {Santos}}{{Pace}
  et~al.}{2009}]{2009A&A...499L...9P}
{Pace} G.,  {Melendez} J.,  {Pasquini} L.,  {Carraro} G.,  {Danziger} J.,
  {Fran{\c c}ois} P.,  {Matteucci} F.,    {Santos} N.~C.,  2009, A\&A, 499, L9

\bibitem[\protect\citeauthoryear{{Pettersen}}{{Pettersen}}{1991}]{1991MmSAI..6%
2..217P}
{Pettersen} B.~R.,  1991, Memorie della Societa Astronomica Italiana, 62, 217

\bibitem[\protect\citeauthoryear{{Pizzolato}, {Maggio}, {Micela}, {Sciortino}
  \& {Ventura}}{{Pizzolato} et~al.}{2003}]{2003A&A...397..147P}
{Pizzolato} N.,  {Maggio} A.,  {Micela} G.,  {Sciortino} S.,    {Ventura} P.,
  2003, A\&A, 397, 147

\bibitem[\protect\citeauthoryear{{Pojmanski}}{{Pojmanski}}{2003}]{2003AcA....5%
3..341P}
{Pojmanski} G.,  2003, Acta Astronomica, 53, 341

\bibitem[\protect\citeauthoryear{{Robinson}, {Cram} \& {Giampapa}}{{Robinson}
  et~al.}{1990}]{1990ApJS...74..891R}
{Robinson} R.~D.,  {Cram} L.~E.,    {Giampapa} M.~S.,  1990, ApJ Suppl., 74,
  891

\bibitem[\protect\citeauthoryear{{Rutten}, {Schrijver}, {Lemmens} \&
  {Zwaan}}{{Rutten} et~al.}{1991}]{1991A&A...252..203R}
{Rutten} R.~G.~M.,  {Schrijver} C.~J.,  {Lemmens} A.~F.~P.,    {Zwaan} C.,
  1991, A\&A, 252, 203

\bibitem[\protect\citeauthoryear{{Rutten}, {Zwaan}, {Schrijver}, {Duncan} \&
  {Mewe}}{{Rutten} et~al.}{1989}]{1989A&A...219..239R}
{Rutten} R.~G.~M.,  {Zwaan} C.,  {Schrijver} C.~J.,  {Duncan} D.~K.,    {Mewe}
  R.,  1989, A\&A, 219, 239

\bibitem[\protect\citeauthoryear{{Saar}}{{Saar}}{2001}]{2001ASPC..223..292S}
{Saar} S.~H.,  2001, in {R.~J.~Garc\'ia Lopez, R.~Rebolo, \& M.~R.~Zapaterio
  Osorio} ed., 11th Cambridge Workshop on Cool Stars, Stellar Systems and the
  Sun Vol.~223 of ASP Conf. Ser., {Recent Measurements of (and Inferences
  About) Magnetic Fields on K and M Stars (CD-ROM Directory: contribs/saar1)}.
p.~292

\bibitem[\protect\citeauthoryear{{Saar} \& {Donahue}}{{Saar} \&
  {Donahue}}{1997}]{1997ApJ...485..319S}
{Saar} S.~H.,  {Donahue} R.~A.,  1997, ApJ, 485, 319

\bibitem[\protect\citeauthoryear{{Schmitt}, {Fleming} \& {Giampapa}}{{Schmitt}
  et~al.}{1995}]{1995ApJ...450..392S}
{Schmitt} J.~H.~M.~M.,  {Fleming} T.~A.,    {Giampapa} M.~S.,  1995, ApJ, 450,
  392

\bibitem[\protect\citeauthoryear{{Schrijver}}{{Schrijver}}{1987}]{1987A&A...17%
2..111S}
{Schrijver} C.~J.,  1987, A\&A, 172, 111

\bibitem[\protect\citeauthoryear{{Schrijver}, {Cote}, {Zwaan} \&
  {Saar}}{{Schrijver} et~al.}{1989}]{1989ApJ...337..964S}
{Schrijver} C.~J.,  {Cote} J.,  {Zwaan} C.,    {Saar} S.~H.,  1989, ApJ, 337,
  964

\bibitem[\protect\citeauthoryear{{Schrijver}, {Dobson} \& {Radick}}{{Schrijver}
  et~al.}{1992}]{1992A&A...258..432S}
{Schrijver} C.~J.,  {Dobson} A.~K.,    {Radick} R.~R.,  1992, A\&A, 258, 432

\bibitem[\protect\citeauthoryear{{Schrijver} \& {Rutten}}{{Schrijver} \&
  {Rutten}}{1987}]{1987A&A...177..143S}
{Schrijver} C.~J.,  {Rutten} R.~G.~M.,  1987, A\&A, 177, 143

\bibitem[\protect\citeauthoryear{{Schrijver} \& {Zwaan}}{{Schrijver} \&
  {Zwaan}}{1991}]{1991A&A...251..183S}
{Schrijver} C.~J.,  {Zwaan} C.,  1991, A\&A, 251, 183

\bibitem[\protect\citeauthoryear{{Schrijver} \& {Zwaan}}{{Schrijver} \&
  {Zwaan}}{2000}]{2000ssma.book.....S}
{Schrijver} C.~J.,  {Zwaan} C.,  2000, {Solar and Stellar Magnetic Activity}.
Cambridge University Press. Cambridge astrophysics series; 34

\bibitem[\protect\citeauthoryear{{Strassmeier}, {Bartus}, {Cutispoto} \&
  {Rodon\'o}}{{Strassmeier} et~al.}{1997}]{1997A&AS..125...11S}
{Strassmeier} K.~G.,  {Bartus} J.,  {Cutispoto} G.,    {Rodon\'o} M.,  1997,
  A\&A Suppl., 125, 11

\bibitem[\protect\citeauthoryear{{Strassmeier}, {Fekel}, {Bopp}, {Dempsey} \&
  {Henry}}{{Strassmeier} et~al.}{1990}]{1990ApJS...72..191S}
{Strassmeier} K.~G.,  {Fekel} F.~C.,  {Bopp} B.~W.,  {Dempsey} R.~C.,
  {Henry} G.~W.,  1990, ApJ Suppl., 72, 191

\bibitem[\protect\citeauthoryear{{Thatcher} \& {Robinson}}{{Thatcher} \&
  {Robinson}}{1993}]{1993MNRAS.262....1T}
{Thatcher} J.~D.,  {Robinson} R.~D.,  1993, Monthly Notices of the RAS, 262, 1

\bibitem[\protect\citeauthoryear{{Torres}, {Quast}, {Melo} \&
  {Sterzik}}{{Torres} et~al.}{2008}]{2008hsf2.book..757T}
{Torres} C.~A.~O.,  {Quast} G.~R.,  {Melo} C.~H.~F.,    {Sterzik} M.~F.,  2008,
  {Young Nearby Loose Associations}.
Handbook of Star Forming Regions, Volume II: The Southern Sky ASP Monograph
  Publications, Vol. 5, p.~757

\bibitem[\protect\citeauthoryear{{Vaughan} \& {Preston}}{{Vaughan} \&
  {Preston}}{1980}]{1980PASP...92..385V}
{Vaughan} A.~H.,  {Preston} G.~W.,  1980, Publications of the ASP, 92, 385

\bibitem[\protect\citeauthoryear{{Walkowicz}, {Hawley} \& {West}}{{Walkowicz}
  et~al.}{2004}]{2004PASP..116.1105W}
{Walkowicz} L.~M.,  {Hawley} S.~L.,    {West} A.~A.,  2004, Publications of the
  ASP, 116, 1105

\bibitem[\protect\citeauthoryear{{West} \& {Hawley}}{{West} \&
  {Hawley}}{2008}]{2008PASP..120.1161W}
{West} A.~A.,  {Hawley} S.~L.,  2008, Publications of the ASP, 120, 1161

\bibitem[\protect\citeauthoryear{{West}, {Hawley}, {Bochanski}, {Covey},
  {Reid}, {Dhital}, {Hilton} \& {Masuda}}{{West}
  et~al.}{2008}]{2008AJ....135..785W}
{West} A.~A.,  {Hawley} S.~L.,  {Bochanski} J.~J.,  {Covey} K.~R.,  {Reid}
  I.~N.,  {Dhital} S.,  {Hilton} E.~J.,    {Masuda} M.,  2008, AJ, 135, 785

\bibitem[\protect\citeauthoryear{{West}, {Hawley}, {Walkowicz}, {Covey},
  {Silvestri}, {Raymond}, {Harris}, {Munn}, {McGehee}, {Ivezi{\'c}} \&
  {Brinkmann}}{{West} et~al.}{2004}]{2004AJ....128..426W}
{West} A.~A.,  {Hawley} S.~L.,  {Walkowicz} L.~M.,  {Covey} K.~R.,  {Silvestri}
  N.~M.,  {Raymond} S.~N.,  {Harris} H.~C.,  {Munn} J.~A.,  {McGehee} P.~M.,
  {Ivezi{\'c}} {\v Z}.,    {Brinkmann} J.,  2004, AJ, 128, 426

\bibitem[\protect\citeauthoryear{{West}, {Walkowicz} \& {Hawley}}{{West}
  et~al.}{2005}]{2005PASP..117..706W}
{West} A.~A.,  {Walkowicz} L.~M.,    {Hawley} S.~L.,  2005, Publications of the
  ASP, 117, 706

\bibitem[\protect\citeauthoryear{{Zuckerman} \& {Song}}{{Zuckerman} \&
  {Song}}{2004}]{2004ARA&A..42..685Z}
{Zuckerman} B.,  {Song} I.,  2004, A\&A Annual Review, 42, 685

\end{thebibliography}
